\documentclass{aa}

\usepackage[english]{babel}
\usepackage[utf8]{inputenc}
\usepackage{amsmath}
\usepackage{amssymb,amsfonts,textcomp}
\usepackage{array}
\usepackage{supertabular}
\usepackage{hhline}
\usepackage{hyperref}
\usepackage[usenames]{color}
\hypersetup{dvips, colorlinks=true, linkcolor=black, citecolor=black, filecolor=blue, urlcolor=blue}
\usepackage{graphicx}
\bibpunct{(}{)}{;}{a}{}{,}

\def\gtrsim{\lower.5ex\hbox{$\; \buildrel > \frac \sim \;$}}

\begin{document}

\title{Cosmic-ray propagation in the bi-stable interstellar medium}
\subtitle{I. Conditions for cosmic-ray trapping}

\titlerunning{CR diffusion in the ISM}
\authorrunning{B. Commer\c con, A. Marcowith, Y. Dubois}

\author{Beno\^it Commer\c con\inst{1}, Alexandre Marcowith\inst{2}, Yohan Dubois\inst{3}}
\institute{ Univ Lyon, Ens de Lyon, Univ Lyon1, CNRS, Centre de Recherche Astrophysique de Lyon UMR5574, F-69007, Lyon, France \\ 
\email{benoit.commercon@ens-lyon.fr} \and Laboratoire Univers et Particules de Montpellier (LUPM) Universit{\'e} Montpellier, CNRS/IN2P3, CC72, place Eug{\`e}ne Bataillon, 34095, Montpellier Cedex 5, France \\\email{Alexandre.Marcowith@umontpellier.fr} \and Institut d'Astrophysique de Paris, UMR 7095, CNRS, UPMC Univ. Paris VI, 98 bis boulevard Arago, 75014 Paris, France \\\email{dubois@iap.fr}}
\date{Received / Accepted }

\abstract
{Cosmic rays propagate through the galactic scales down to the smaller scales at which stars form. Cosmic rays are close to energy equipartition with the other components of the interstellar medium and can provide a support against gravity if pressure gradients develop.}
{We study the propagation of cosmic rays within the turbulent and magnetised bi-stable interstellar gas. The conditions necessary for cosmic-ray trapping and cosmic-ray pressure gradient development are investigated.}
{We derived an analytical value of the critical diffusion coefficient for cosmic-ray trapping within a turbulent medium, which follows the observed scaling relations. 
We then presented a numerical study using 3D simulations of the evolution of a mixture of interstellar gas and cosmic rays,  in which turbulence is driven at varying scales by stochastic forcing within a box of 40~pc. We explored a large parameter space in which the cosmic-ray diffusion coefficient, the magnetisation, the driving scale, and the amplitude of the turbulence forcing, as well as the initial cosmic-ray energy density, vary.}
{ We identify a clear transition in the interstellar dynamics for cosmic-ray diffusion coefficients below a critical value deduced from observed scaling relations. This critical diffusion depends on the characteristic length scale $L$ of $D_{\rm crit}\simeq 3.1\times 10^{23}$~cm$^2$~s$^{-1} (L/`\rm{1~pc})^{q+1}$, where the exponent $q$ relates the turbulent velocity dispersion $\sigma$ to the length scale as $\sigma\sim L^q$.  Hence, in our simulations this transition occurs around $D_{\rm crit} \simeq 10^{24}-10^{25}$~cm$^2$~s$^{-1}$. The transition is recovered in all cases of our parameter study and is in very good agreement with our simple analytical estimate. In the trapped cosmic-ray regime, the induced cosmic-ray pressure gradients can modify the gas flow and provide a support against the thermal instability development. We discuss possible mechanisms that can significantly reduce the cosmic-ray diffusion coefficients within the interstellar medium.}
{Cosmic-ray pressure gradients can develop and modify the evolution of thermally bi-stable gas for diffusion coefficients $D_{0} \leq 10^{25}$~cm$^2$~s$^{-1}$ or in regions where the cosmic-ray pressure exceeds the thermal one by more than a factor of ten. This study provides the basis for further works including more realistic cosmic-ray diffusion coefficients, as well as local cosmic-ray sources.}


\maketitle


\section{Introduction}

Cosmic rays (CRs) are an important component of the interstellar medium (ISM); their pressure is close to equipartition with magnetic, gas, and radiation pressures, all of the order of 1~eV~cm$^{-3}$ in the local ISM \citep{grenier:15}. The details of CR propagation in the ISM are still not well understood. CRs propagate partly by a succession of random walks and their mean free path between two scatterings is poorly known. Therefore the process is controlled by the properties of local magnetic turbulence, which is poorly modelled. Furthermore, while pervading the ISM CRs can generate their own turbulence. This is one of the main ways they back-react over ISM turbulence and its dynamics \citep{kulsrud:69}.

CRs have a multi-scale impact over galactic ISM dynamics \citep{grenier:15}:  they drive large-scale winds \citep{hanasz:13, booth:13, salem:14, girichidis:16, wiener:17}, generate large-scale magnetic fields through the Parker instability \citep{hanasz:09, butsky:17}, inject turbulent magnetic motions at intermediate scales \citep{rogachevskii:12}, contribute to ionise, diffuse, and heat molecular clouds or protostellar environments \citep{padovani:09, rodgers:17, padovani:18}, and finally they produce radio elements by spallation reaction in young stellar systems \citep{ceccarelli:14}. There, CRs as charged fast moving particles have an impact on dust charge  \citep{Ivlev15}.

ISM turbulence is a key mechanism for models of star formation rate and stellar initial mass function \citep[e.g.][]{hennebelle:08,hennebelle:11, padoan:11, federrath:12}. ISM turbulence is characterised by its statistical properties; in particular, as turbulent motions in the ISM are often supersonic, they can be characterised by the density variance-Mach number relation \citep{vazquez:94, krumholz:05, kritsuk:07}. In studies to date, several physical effects have been included to investigate this relation: effects due to different turbulent forcing geometries \citep[compressive versus solenoidal modes, ][]{federrath:10}, the impact of magnetic fields \citep{lemaster:08, molina:12},  and thermodynamic properties of the turbulence in non-isothermal turbulent models  \citep[e.g.][]{nolan:15}. It has been found that the turbulent forcing geometries and the magnetic fields affect the width of the log-normal distribution that fits the probability density function (PDF). In addition to the above-mentioned effects, CRs exert a force over the plasma along their pressure gradients. Hence, this force can also modify ISM dynamics and statistics. These effects should be expected to be particularly strong close to CR sources like supernova remnants or massive star clusters, over spatial scales and timescales that are strongly dependent on the ambient ISM phases surrounding the sources \citep{nava:16}.

One of the goals of this study is to test the impact of anisotropic diffusion of CRs over the gas probability density function (PDF). It should be noticed that at smaller galactic scales corresponding to molecular clouds, little is known about the potential effect of CR gradients on the collapse of molecular clouds \citep[see however][]{everett:11}, especially again if the cloud is near a CR source. This aspect will be treated in a forthcoming and associated study. In this study we first address the conditions in which the force due to a CR gradient can modify ISM turbulent statistics. We compute the response of a bi-stable fluid to different CR diffusivity laws. An associated study will investigate how CRs can modify this diffusivity through the generation of magnetic fluctuations and consider the effect of CR sources.\\

The lay-out of the paper is as follows. In Sect. \ref{section:analysis}, we discuss the necessary conditions on the CR spatial diffusion coefficient to produce a modification of ISM dynamics through the parallel CR gradient. In Sect. \ref{section:numerics} we detail the simulation designed in this study. The main results of our parametric numerical study are presented in Sect.~\ref{section:results}. We discuss our results and their limitations in Sect.~\ref{section:discussion}, and we propose a list of different mechanisms that could lead to variations of the CR diffusion coefficient in the ISM. We conclude and present some perspectives on this study in Sect.~\ref{section:conclusion}.

\section{Qualitative analysis of the properties of the turbulent ISM including cosmic rays}\label{section:analysis}

\subsection{Turbulence and magnetic fields in the ISM}

In this section, we look at the conditions under which CRs can be trapped within the different phases of the ISM. We focus in this short analytical study on the densest phases of the ISM:  the warm neutral medium (WNM) and the cold neutral medium (CNM), essentially composed of neutral atomic hydrogen (HI). Dense ISM is also composed of   molecular clouds, essentially composed of molecular hydrogen (H$_2$), which we also discuss qualitatively, but which we do not investigate using numerical experiments in the present study.  Numerical models of molecular cloud formation and evolution will be the subject of a future study. This analysis is qualitative and is not aimed at covering the wide range of properties of the different gas phases in the ISM.

\subsubsection{Turbulence}\label{section:turbulent}

Turbulence is observed in the ISM at various scales and appears to be universal \citep{larson:81,heyer:04}. 
Recent observations by \cite{burkhart:15} report on transonic turbulence from HI emission PDF, which traces both the CNM and the WNM, while the HI absorption lines from the CNM show supersonic Mach numbers with a median value of 4. 

Empiric relations between the gaseous velocity dispersion $\sigma$ and the region size $L$ are observed in the ISM, which are suggestive of a turbulent cascade \citep{larson:79,larson:81,miville:17}. The turbulent velocity dispersion increases with the size as
\begin{equation}
\sigma \simeq \sigma_\mathrm{1~pc} L^q,
\end{equation}
where $\sigma_\mathrm{1~pc} $ is the velocity dispersion at a scale of 1~pc. 
For a Kolmogorov law, we have $q=1/3$. 

In HI clouds, made of a mixture of WNM and CNM, observations report values of $q\simeq 0.35$ with  $\sigma_\mathrm{1~pc} \simeq 1-1.5$~km~s$^{-1}$ \citep{larson:79,roy:08}, which is close to the expected value for a Kolmogorov cascade.
In the densest part of CNM, the molecular clouds, it is currently accepted that the gas follows scaling relations, named the Larson relations \citep{larson:81}, which relate the non-thermal velocity fluctuations to the size as well as to the mass of molecular clouds.
Following the scaling given in \cite{hennebelle:12}, the Larson relations read 
\begin{eqnarray}\label{Eq:LAR}
  \sigma &\simeq &1 \mathrm{~km~s}^{-1} \left( \frac{L}{\mathrm{1~pc}}\right)^{0.5},\label{eq:Larson_v}\\
   n &\simeq &3000 \mathrm{~cm}^{-3} \left( \frac{L}{\mathrm{1~pc}}\right)^{-0.7},\label{eq:Larson_dens}
\end{eqnarray}
where $n$ is the gas number density, and $L$ is the size of the region in which the non-thermal velocity and the density are measured. It is generally assumed that this length scale represents the molecular cloud size. In this case, it follows that
\begin{equation}
\sigma \simeq 1 \mathrm{~km~s}^{-1} \left( \frac{n}{3000 \mathrm{~cm}^{-3}}\right)^{-5/7}.
\end{equation}

These scaling relations indicate that velocity fluctuations are reduced as the gas becomes denser, until self-gravity takes over to initiate the gravitational protostellar collapse and to re-accelerate the gas on much smaller scales. We do not consider molecular cloud and collapsing dense core scales in our numercial work. Hereafter, we assume that non-thermal motions are due to the turbulence only and consider that the non-thermal velocity is the rms velocity dispersion $v_\mathrm{rms}$.

It is worth noting that the exponent values $q\simeq 0.35-0.5$ in the velocity-size relation are commonly retrieved in numerical works  \citep[e.g.][]{audit:10,saury:14}. In the following, we assume that the velocity dispersion and the size scale as
\begin{equation}
v_\mathrm{rms} \simeq 1 \mathrm{~km~s}^{-1} \left( \frac{L}{\mathrm{1~pc}}\right)^{q},\label{eq:vL}
\end{equation}
with $q$ varying in the different phases of the ISM.

\subsubsection{Magnetic fields}

Magnetic fields are observed at all scales in the ISM and show typical variations as a function of the density. Similarly to the turbulence, magnetic fields induce processes that are highly anisotropic. Magnetic fields are not amplified if the motion occurs along the field lines.  In contrast, if a cloud is contracting, the anisotropic magnetic force will channel the gas to collapse first along the mean magnetic field direction, and then perpendicular to it. For a spherical contraction, the conservation of magnetic flux $BR^2$ for an enclosed mass $M\propto nR^3$ implies that $B\propto n^{2/3}$. The mass-to-flux conservation gives $B\propto n h$, where $h$ is the thickness of the cloud along the field lines. Equilibrium along the field lines implies that $c_\mathrm{s}^2 \sim \phi$, with $c_\mathrm{s}$ the sound speed and $\phi$ the gravitational potential. Finally, the Poisson equation leads to $\phi\propto c_\mathrm{s} h^2$. We thus get $B\propto n^{1/2}$ for an isothermal medium.

Measurements of the magnetic intensity via the Zeeman effect report on two different regimes for the magnetic intensity-density scaling \citep[e.g.][]{crutcher:12}. At low density $n<300$~cm$^{-3}$, the magnetic intensity is insensitive to the density variations, with $B_{0,\mathrm{max}}\simeq 10~\mu$G. While for densities $n>300$~cm$^{-3}$, that is within molecular clouds,  the magnetic intensity scales with the density with an upper value given by $B_\mathrm{max}\propto n^{0.65}$ (which would correspond to a spherical contraction). 

We can thus infer a simple relation that gives the magnetic intensity as a function of the density 
\begin{equation}
B\leq 10~\mu\mathrm{G}~ \left[1+\left(\frac{n}{300~\mathrm{cm}^{-3}}\right)\right]^{0.65}.
\end{equation}

This relation is purely empirical and shows the most probable maximum values of the magnetic field amplitude \citep{crutcher:12}. It does not reflect the variations of the  magnetic amplitude within the ISM, but it is useful to illustrate the different regimes that are expected in the ISM magneto-hydrodynamics (MHD) turbulence. 

\subsubsection{MHD turbulence}

We can derive two regimes for the MHD turbulence, which depends on the Alfv\'{e}nic Mach number $\mathcal{M}_\mathrm{A}$, defined as $\mathcal{M}_\mathrm{A}=v_\mathrm{rms}/v_\mathrm{A}$, with $v_\mathrm{A}=B/(4\pi\mu_\mathrm{gas}m_\mathrm{H}n)^{0.5}$ the Alfv\'en speed and $\mu_\mathrm{gas}$ the gas mean molecular weight. By combining the above-mentioned scaling relations for $B$ and $v_\mathrm{rms}$ at the densities $n<100$~cm$^{-3}$ we interest us in this study, the Alfv\'{e}nic Mach number can be estimated as
\begin{equation}
\mathcal{M}_\mathrm{A}\simeq 0.9 \left(\frac{L}{\mathrm{1~pc}}\right)^{0.35}\left(\frac{n}{100~\mathrm{cm}^{-3}}\right)^{0.5},
\label{Ma_LB}
\end{equation}
where we have taken a constant magnetic field amplitude of 6~$\mu$G according to the results of \cite{heiles:05} for the CNM. We further assume that this value is also valid for the WNM. The above relation only applies for the CNM and WNM scaling, that is, for  $q=0.35$.

\begin{figure}[t]
        \includegraphics[width=0.5\textwidth]{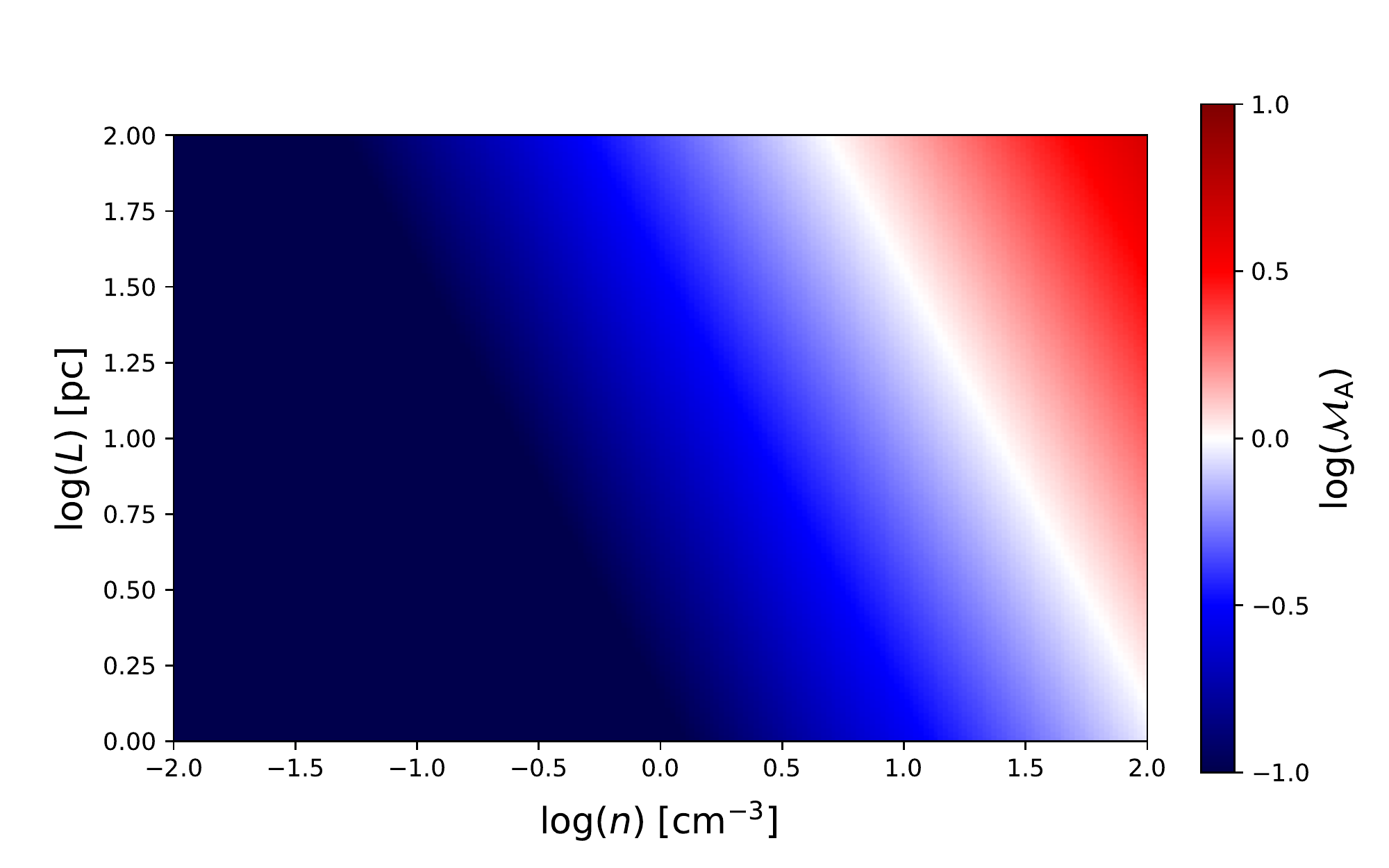}
        \caption{Alfv\'{e}nic Mach number $\mathcal{M}_\mathrm{A}$ as a function of the gas density $n$ and of the length scale $L$ following Eq.~(\ref{Ma_LB}). This plot is valid only if one assumes the scaling relation and magnetic field strength observed in the CNM and WNM.}
        \label{Fig:Ma_MC}
\end{figure}

Figure \ref{Fig:Ma_MC} illustrates the value of the Alfv\'{e}nic Mach number $\mathcal{M}_\mathrm{A}$ as function of the gas density and the length scale from Eq.~\ref{Ma_LB}. This figure is only shown as an illustrative example of the range of  Alfv\'{e}nic Mach numbers that are observed (see Fig.~\ref{Fig:k2_scatter} for instance).
The MHD turbulence within the ISM switches between the regime of super- and sub-Alfv\'{e}nic turbulence if the flow satisfies the above-mentioned scaling relations. At a gas density of $\leq 10$~cm$^{-3}$, the flow is always sub-Alfv\'{e}nic, while at larger densities it may be super-Alfv\'{e}nic. CRs propagate along magnetic field lines, and the topology of the lines thus affects the global propagation speed. In the super-Alfv\'{e}nic regime, the magnetic field lines are tangled and the CRs take some time to escape from turbulent eddies. This reduces the effective diffusion coefficient. In addition, the turbulence tends to isotropize the diffusion process so that the effective parallel and perpendicular coefficients are equal. On the contrary, CRs can propagate rapidly along smooth magnetic field lines in the sub-Alfv\'{e}nic regime. The parallel diffusion coefficient is large, but the perpendicular diffusion coefficient is typically orders of magnitude smaller \citep[e.g.][]{yan:2004}.

\subsection{The critical CR diffusion coefficient\label{sec:Dcrit}}

We consider a blob of turbulent interstellar gas of length $L$, with a turbulence characterised by the root-mean-square velocity $v_\mathrm{rms}$. The turbulent time is defined as the time at which a turbulent fluctuation travels through the blob,
\begin{equation}
  t_\mathrm{turb}\equiv\frac{L}{v_\mathrm{rms}}.
  \label{eq:tturb}
\end{equation}
We define the diffusion time $t_\mathrm{diff}$ as
\begin{equation}
t_\mathrm{diff}\equiv\frac{L^2}{D},
\end{equation}
where $D$ is the CR diffusion coefficient.

It is straightforward to estimate a critical diffusion coefficient $D_\mathrm{crit}$ that corresponds to the regime where CR diffusion is not sufficient to prevent energy fluctuations. CRs then get trapped, which can lead to a significant gradient of CRs and, thus, a dynamical effect of the CR pressure onto the flow. 

The critical diffusion is given at $t_\mathrm{turb}=t_\mathrm{diff}$
\begin{equation}
D_\mathrm{crit}\equiv L v_\mathrm{rms}. 
\label{Eq:defDcrit}
\end{equation}

Using the scaling relation in Eq.~(\ref{eq:vL}), the critical diffusion coefficient is
\begin{equation}
D_\mathrm{crit} \simeq 3.1 \times 10^{23} \mathrm{~cm}^2\mathrm{~s}^{-1} \left(\frac{L}{\mathrm{1~pc}}\right)^{q+1}.
\label{Eq:Dcrit}
\end{equation}

The above calculation of $D_{\rm crit}$ is an order of estimate as the scaling relation for $v_{\rm rms}$ shows some dispersion and as we neglect the dimensionality of the diffusion process. Nevertheless, the value of $q$ does not vary a lot between the WNM and CNM and the molecular clouds so that our prediction is robust and does not significantly change in the different phases of the ISM. 
For a cloud 10 pc in size, which is typical of the length used in studies of molecular clouds turbulence \citep[e.g.][]{federrath:12}, the critical diffusion coefficient is $D_\mathrm{crit} \simeq 10^{25}$ cm$^{2}$~s$^{-1}$.  
A usual parametrisation of the CR parallel diffusion coefficient derived from CR data fitting is 
\begin{equation}
D_\mathrm{ISM}\sim 10^{28-29}\mathrm{~cm}^2\mathrm{~s}^{-1}\left(\frac{E}{1~\rm Gev}\right)^s,
\end{equation}
with $s\simeq 0.3-0.6$ \citep{strong:07}. 
The value of $D_\mathrm{crit}$ we derived is small compared to the canonical value of  $D_\mathrm{ISM}\simeq 10^{28}$~ cm$^{2}$~s$^{-1}$  at a kinetic energy of $\simeq$ 1 GeV \cite[e.g.][]{ptuskin:06} in which we are interested in this study. CR energy spectra peak around $\simeq$ 1 GeV in Galactic environments \citep{grenier:15}, which thus corresponds to the CR particles that contribute most to the CR pressure. We discuss in Sect. \ref{section:discussion} some possible mechanisms that can modify the CR diffusion coefficient in the different phases of the ISM.

\section{Numerical setup}
\label{section:numerics}

\begin{table*}
        \caption{Summary of the initial conditions of all the simulations. From left to right columns: Run name; ratio of CR to thermal pressure; CR parallel diffusion coefficient; turbulence driving wave number in units of the box length; turbulent forcing velocity; turbulence crossing time at the box scale length $t_\mathrm{cross}=L_\mathrm{box}/v_\mathrm{rms}$; Alfv\'enic Mach number; critical diffusion coefficient defined as in Eq.~(\ref{eq:Dcrit_simu}).}
        \centering
        \begin{tabular}{cccccccc}
                \hline
                \hline
                Model     &$\zeta=P_\mathrm{CR}/P$      & $D_0$                   & $k_\mathrm{turb}$  & $v_\mathrm{rms}$ & $t_\mathrm{cross}$ & $\mathcal{M}_\mathrm{A}$ & $D_\mathrm{crit}$\\
                          &                                             & [cm$^2$ s$^{-1}$]   &                                           & [km s$^{-1}$] & [Myr]                                 &  & [cm$^2$ s$^{-1}$]
                          \\
                \hline
                \vspace{2pt}
                k2ani     & 0.1                 & $0-10^{22}-10^{24}-10^{26}-10^{28}$ &   2       &   5.7                       &     6.9             & 4.4   & $3.2\times10^{25}$\\            
                k4ani     & 0.1                 & $0-10^{22}-10^{24}-10^{26}-10^{28}$ &   4       &   3.9                       &     10              & 3             &  $1.2\times10^{25}$\\           
                k8ani     & 0.1                 & $0-10^{22}-10^{24}-10^{26}-10^{28}$ &   8       &   2.9                       &     13.5    & 2.2   &        $4.4\times10^{24}$\\            
                k16ani    & 0.1                 & $0-10^{22}-10^{24}-10^{26}-10^{28}$ &   16      &   1.8                       &     21.7    & 1.4&   $1.35\times10^{24}$    \\              
                \\
                NoCRk$X$\tablefootmark{a}& 0    & -                                &   $2-4-8-16$   &                &   & 4.4   & \\    
                \\
                bk2ani    &     $1-10-100$              & $10^{24}-10^{26}$ &   2       &           5.7               &     6.9             & 4.4& $3.2\times10^{25}$       \\              
                \\              
                k2rmsx2   & 0.1                 & $10^{22}-10^{24}-10^{26}-10^{28}$   &   2     &   10.5                      &     3.7     & 8.2 & $6.3\times10^{25}$              \\              
                \\
                Mak2ani   & 0.1                 & $10^{24}-10^{26}$   &   2          &   5.7                       &     6.9             & $0.5-1-20$ & $3.2\times10^{25}$    \\              
                \\
                isok2ani  & 0.1                 & $10^{22}-10^{24}-10^{26}-10^{28}$   &   2     &           5.7               &     6.9             & 4.4 & $3.2\times10^{25}$      \\              
                \\
                \hline
        \end{tabular}
        \tablefoot{
                \tablefoottext{a}$X$ stands for the value of the forcing wave number $k_\mathrm{turb}$.
        }
        \label{tab:summary}
\end{table*}

\subsection{Magneto-hydrodynamics with cosmic-ray diffusion}

Our numerical model integrates the equation of ideal MHD for a fluid mixture made of gas and CRs, including anisotropic CR diffusion. We describe in the following the details of our numerical tool. 

We use the adaptive-mesh-refinement code \ttfamily{RAMSES}\rm~\citep{teyssier:02}, which is based on a finite-volume, second-order Godunov scheme and a constrained transport algorithm for ideal MHD \citep{fromang:06,teyssier:06}. We use the HLLD (Harten-Lax-Van Leer-Discontinuities) Riemann solver to compute the MHD flux, as well as the electric field to integrate the induction equation \citep{miyoshi:05}. CRs are treated as a fluid following the implementation of anisotropic diffusion described by \cite{dubois:16}. It consists in an advection-diffusion equation of the CR energy density, in which the CRs are advected with the fluid, contribute to the total energy, and propagate along the magnetic field lines. 
The full CR-MHD equations  read
\begin{eqnarray}
\frac{\partial \rho}{\partial t} &+& \nabla\cdot \left[\rho\vec{u} \right] =  0, \\
\frac{\partial \rho \vec{u}}{\partial t} & + &\nabla \cdot\left[\rho \vec{u} \vec{u} + P_\mathrm{tot} \mathbb{I} - \frac{\vec{B} \vec{B}}{4\pi} \right] = \rho \vec{f} ,\\
\frac{\partial E_\mathrm{tot}}{\partial t} & + &\nabla\cdot \left[\vec{u}\left( E_\mathrm{tot} + P_\mathrm{tot} \right) -\frac{\left(\vec{u}\cdot\vec{B}\right)\vec{B}}{4\pi}\right] \nonumber\\
&=&  - P_\mathrm{CR}\nabla\cdot\vec{u}  + \nabla\cdot\mathbb{D}\nabla E_\mathrm{CR} \\
& & +\rho \vec{f}\cdot\vec{u}  - \mathcal{L}(\rho,T) \nonumber\\
\frac{\partial E_\mathrm{CR}}{\partial t}& + &\nabla\cdot \left[\vec{u}E_\mathrm{CR}\right]
= 
- P_\mathrm{CR}\nabla\cdot\vec{u} + \nabla\cdot\mathbb{D}\nabla E_\mathrm{CR}, \\
\frac{\partial \vec{B}}{\partial t}& -& \nabla\times\left[\vec{u}\times\vec{B}\right] =0,
\end{eqnarray}

\noindent where $\rho$ is the gas density, $\vec{u}$ is the fluid velocity, $\mathbb{I}$ is the identity matrix, $P_\mathrm{tot}$ the total pressure, that is, the sum of the gas thermal pressure $P=(\gamma-1)E$ and the CR pressure $P_\mathrm{CR}=(\gamma_\mathrm{CR}-1)E_\mathrm{CR}$ where $\gamma$ and $\gamma_{\rm CR}$ are respectively the gas and CR adiabatic index (and $E$ and $E_{\rm CR}$ the gas and CR internal energy densities), and the magnetic pressure $P_\mathrm{mag}=B^2/(8\pi)$. The magnetic field vectors is $\vec{B}$ ,  $E_\mathrm{tot}$ is the total energy $E_\mathrm{tot}= E +0.5\rho u^2 + B^2/(8\pi) + E_\mathrm{CR}$. The anisotropic diffusion tensor of cosmic rays is $\mathbb{D}$ , given by
\begin{equation}
\mathbb{D}=D_{\perp} \delta_{ij} - (D_\perp -D_\parallel)\vec b \vec b,
\end{equation}
where $\vec b=\vec B/||\vec B||$, $D_\parallel$ and $D_\perp$ are the diffusion coefficients respectively along and across the magnetic field. Their amplitude remains poorly constrained in the turbulent ISM. For numerical convenience, we set $D_\perp=0.01D_\parallel$ in order to avoid poor convergence of our implicit solver used for the anisotropic diffusion \citep{dubois:16}.  For the thermal budget, we include heating and cooling processes $ \mathcal{L}(\rho,T)$ following \cite{audit:05} to be able to describe the CNM and the WNM phases of the ISM, as well as the thermal instability \citep{field:1965}. The impact of gas heating due to CR streaming remains to be quantified with respect to the classical heating and cooling processes included in this work. We also neglect the role of CR streaming and CR loss processes \citep[see for instance][for alternative implementations of CR that account for some of these effects]{ruszkowski:17,pfrommer:17}. The CR streaming velocity magnitude corresponds to the velocity of the Alfv\'{e}n waves $v_\mathrm{A}$. We thus expect that CR streaming may have an impact in the sub-Alfv\'{e}nic regimes.  In addition, CR streaming leads to a loss term in the CR energy equation, and thus to a heating term for the gas thermal energy.  These two effects will be investigated in future works. Lastly, CR cooling rates in the ionised ISM are often accounted for \citep[e.g.][]{guo:08,booth:13}, but the cooling rates in the neutral ISM remain to be quantified. 

The specific force $\vec{f}$ represents the force that is applied to drive the turbulence. This force field is generated in the Fourier space  and its mode is given by a stochastic process based on the Ornstein-Uhlenbeck process  \citep{eswaran:88,schmidt:06}. For more details on the turbulence forcing module, we refer readers to the work of \cite{schmidt:09} and \cite{federrath:10} on which our implementation is based. We set the forcing to a mixture of compressive and solenoidal modes, with a compressive force power equal to $1/3$ of the total forcing power. We use a unique autocorrelation timescale $T\simeq 0.6$~Myr. We neglect self-gravity in this study. 

The system is closed using the perfect gas relation, $P=\rho k_\mathrm{B}T/(\mu_\mathrm{gas} m_\mathrm{H})$, with $\mu_\mathrm{gas}=1.4$. We set the gas adiabatic index to $\gamma=5/3$ and the CR fluid adiabatic index to $\gamma_\mathrm{CR}=4/3$. Lastly, we define the modified sonic Mach number for the gas and CR mixture as \begin{equation}
\mathcal{M}_\mathrm{s,CR}\equiv\frac{v}{c_\mathrm{s,CR}},
\end{equation} 
where 
\begin{equation}
c_\mathrm{s,CR}\equiv\sqrt{\gamma \frac{P}{\rho}+\gamma_\mathrm{CR} \frac{P_\mathrm{CR}}{\rho}}
\label{Eq:cs}
\end{equation} 
is the modified sound speed which accounts for the CR pressure.

\subsection{Initial conditions}

Our initial set-up is very similar to that of \cite{seifried:11} and \cite{saury:14}. We set a box of size $L_\mathrm{box}=40$~pc filled with gas that follows the observed scaling relations. The initial state of the gas corresponds to a thermally unstable state. The initial temperature is $T_0\simeq4460$~K, density $n_0=2$~cm$^{-3}$, gas thermal pressure $P\simeq1.23\times10^{-12}$~dyne~cm$^{-2}$. The initial sound speed is  $c_{\mathrm{s}}\simeq 5.2 $~km~s$^{-1}$. The initial cosmic-ray pressure is set by its ratio with the gas thermal pressure, which we define as $\zeta \equiv P_\mathrm{CR}/P$.  If not stated otherwise, we set $\zeta=0.1$. The initial magnetic field is set to $B_0=1~\mu$G, which gives an initial Alfv\'{e}nic speed $v_\mathrm{A}\simeq 1.3$~km~s$^{-1}$ and a plasma beta $\beta=P/P_\mathrm{mag}\simeq30$. Magnetic field amplification by turbulent dynamo is unlikely to take place in our models since we start with a magnetic field amplitude that corresponds to the values expected at saturation \citep[e.g.][]{ferriere:00,rieder:18}.

The turbulence is driven with various mode wave numbers $k$ that correspond to a fraction of the box size. For $k_\mathrm{turb}=2$, the turbulence is driven at a scale corresponding to half the box size. We explore four different driving scales, $k_\mathrm{turb}=2$, $k_\mathrm{turb}=4$, $k_\mathrm{turb}=8$, and $k_\mathrm{turb}=16$. This setup mimics the effect of different driving mechanisms such as supernovae at large driving scales and protostellar jets at the smallest scales. We use a parabolic function between $k\in[k_\mathrm{turb}-1,k_\mathrm{turb}+1]$ to choose the wave numbers for the forcing module. The amplitude of the driving is kept constant between all the different scales, which means that the resulting $v_\mathrm{rms}$ varies as a function of $k_\mathrm{turb}$.

\subsection{Parametric study}
\begin{figure}[t]
        \includegraphics[width=0.5\textwidth]{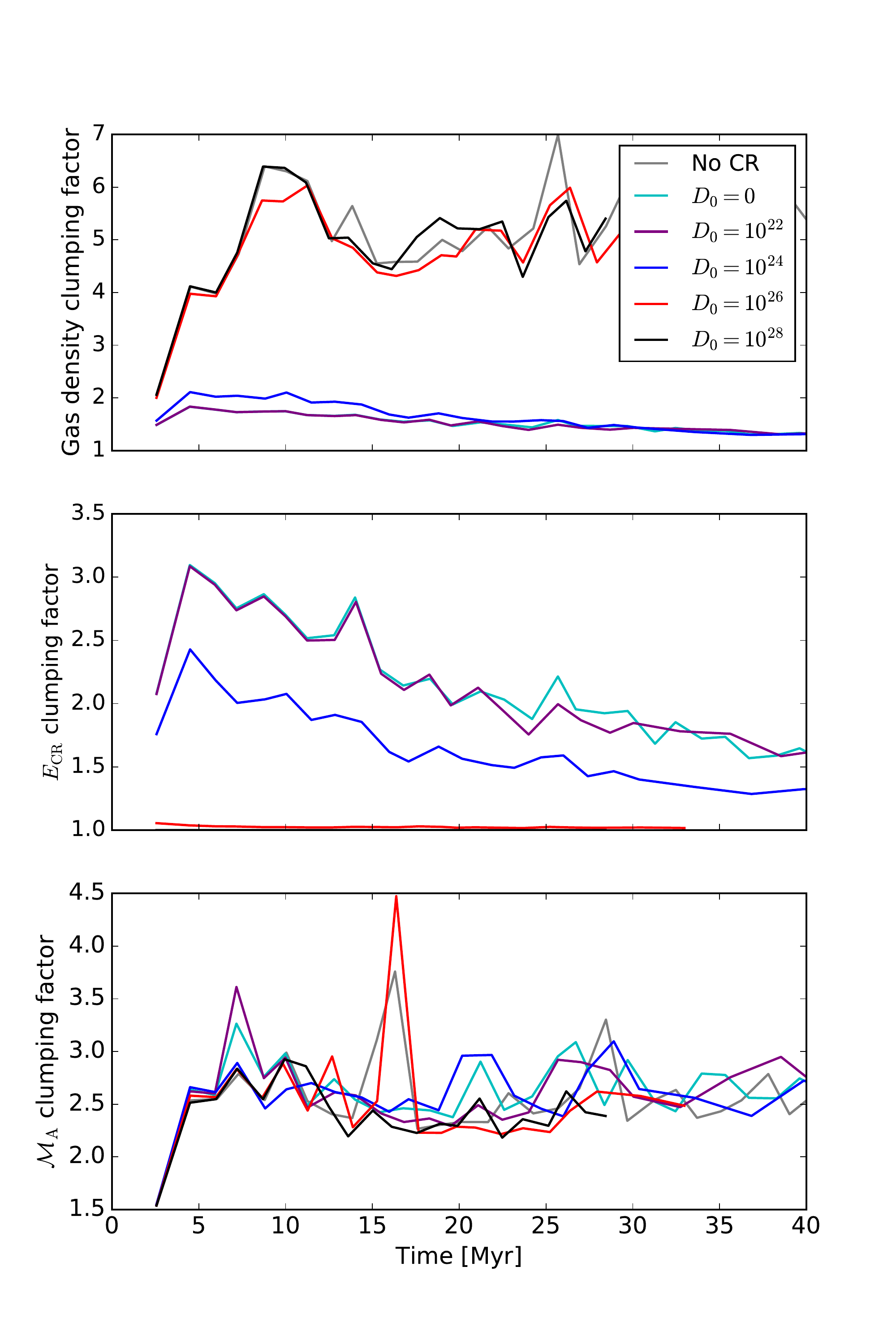}
        \caption{Time evolution for six simulations with $k_\mathrm{turb}=2$ of the clumping factor of the gas density $n$ (top), the cosmic-ray energy density $E_\mathrm{CR}$ (middle), and the Alfv\'enic Mach number $\mathcal{M}_\mathrm{A}$ (bottom). The clumping factor $C(X)$ of a quantity $X$ is defined as $C(X)=<X^2>/<X>^2$.}
        \label{Fig:time_k2}
\end{figure}

\begin{figure*}[!t]
        \centering
        \resizebox{\hsize}{!}{\includegraphics{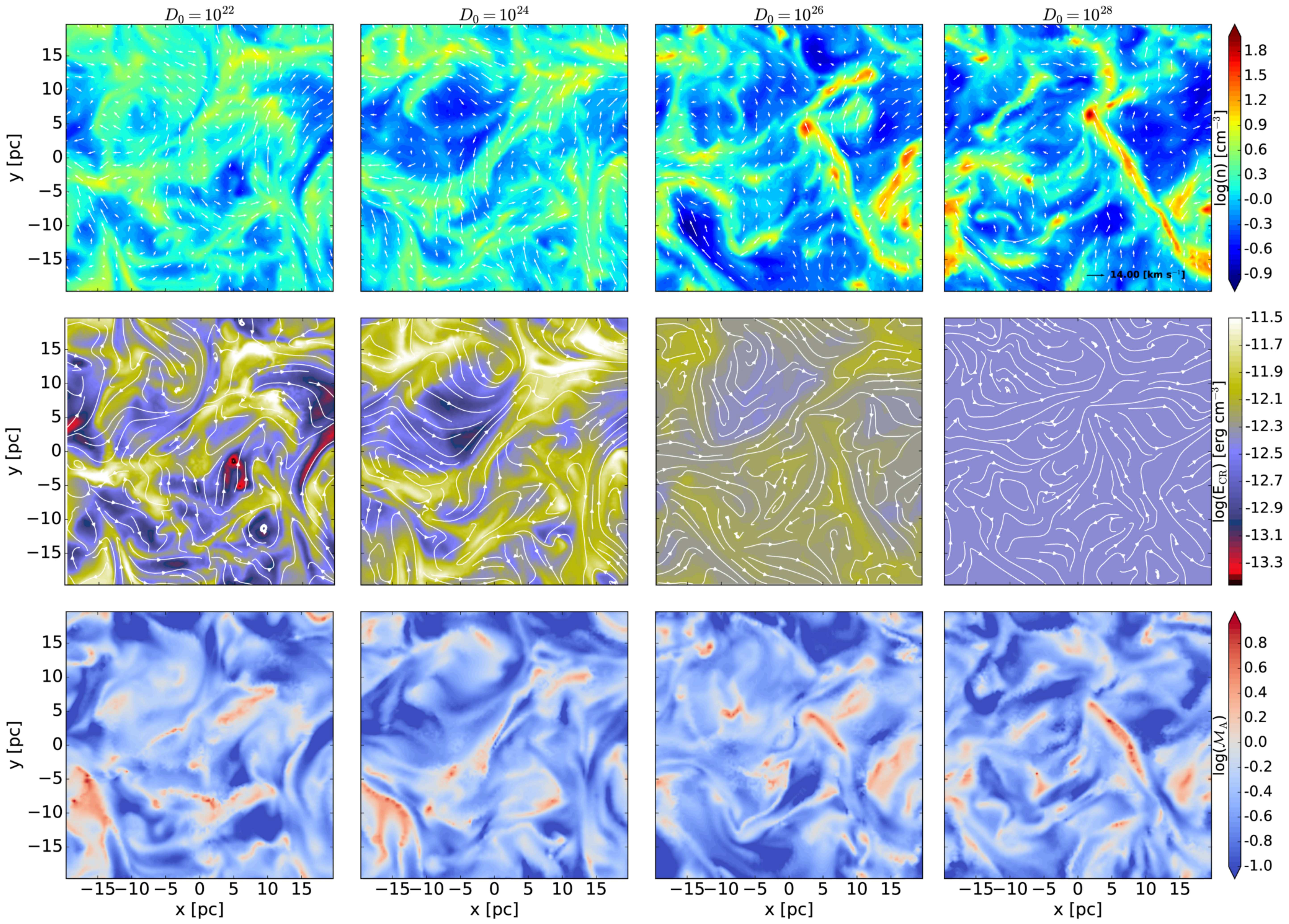}}
        \caption{Fiducial runs with $k_\mathrm{turb}=2$: maps of the gas density (top), cosmic-ray energy density (middle), and Alfv\'enic Mach number (bottom) in the plane $z=20$~pc at a time corresponding to $2t_\mathrm{cross}$. From left to right, the CR diffusion coefficient is $D_0=10^{22}$, $10^{24}$, $10^{26}$, and $10^{28}\, \rm cm^2 \, s^{-1}$. The arrows on the density maps represent the velocity vectors (normalised to a value of 14 km~s$^{-1}$, see upper right panel), and the streamlines on the CR energy density maps represent the magnetic field lines, both in the $xy$-plane.}
        \label{Fig:map_k2}
\end{figure*}
\begin{figure*}[!t]
        \centering
        \resizebox{\hsize}{!}{\includegraphics{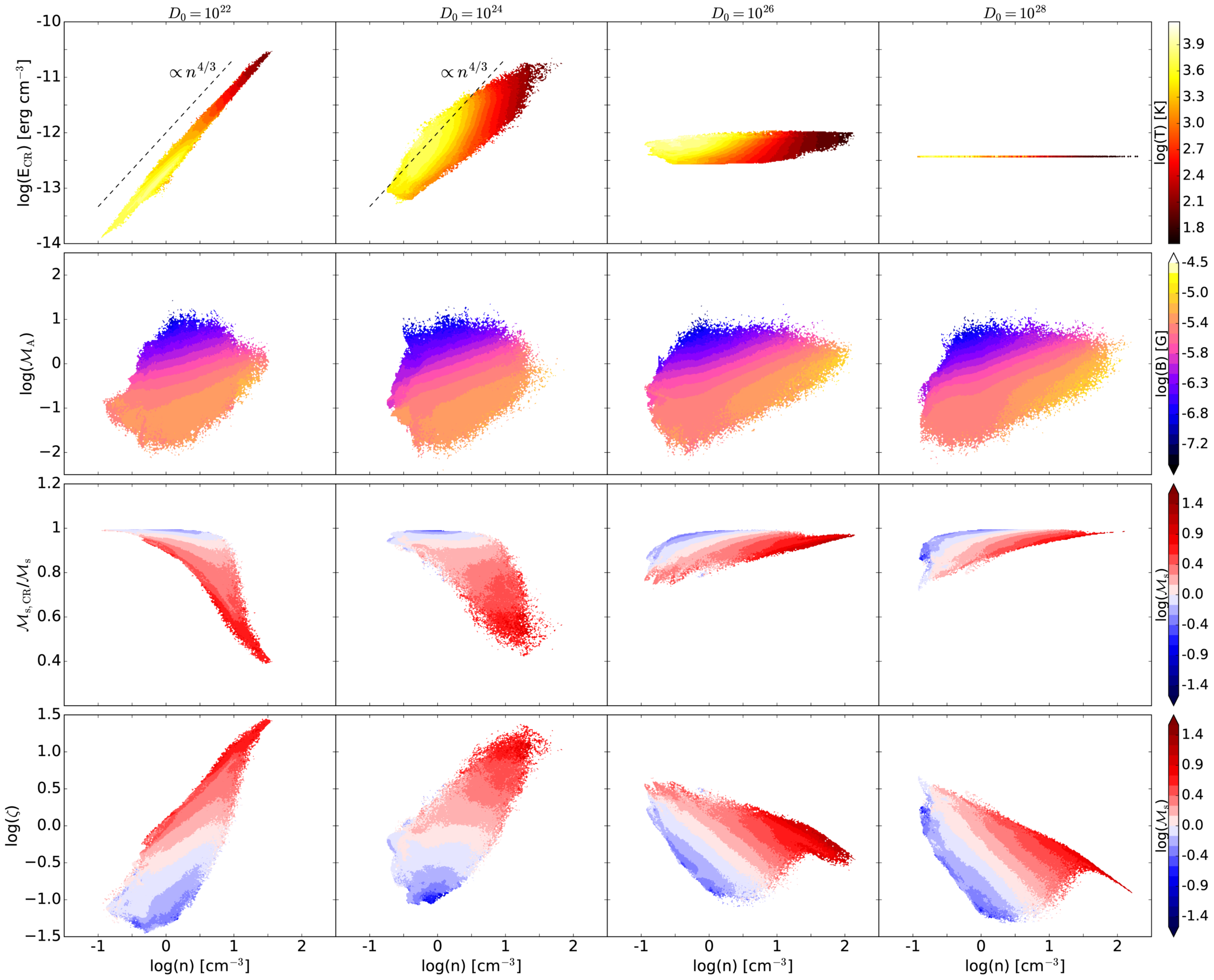}}
        \caption{Two-dimensional histograms of, from top to bottom, the CR energy density, the   Alfv\'enic Mach number $\mathcal{M}_\mathrm{A}$, the ratio $\mathcal{M}_\mathrm{s,CR}/\mathcal{M}_\mathrm{s}$, and $\zeta=P_\mathrm{CR}/P$ as a function of the density  for the fiducial runs $k_\mathrm{turb}=2$ with CR diffusion at the same time as in Fig.~\ref{Fig:map_k2}. The colour coding indicates the temperature, the magnetic field strength, and the sonic Mach number $\mathcal{M}_\mathrm{s}$ (two bottom rows) respectively.}
        \label{Fig:k2_scatter}
\end{figure*}

We performed a set of 50 simulations to explore the initial conditions of the parameter space. We used a uniform fiducial resolution of $128^3$. \cite{saury:14}  performed a resolution study (from $128^3$ to $512^3$) and have shown that even though the scales to describe the cold CNM and the molecular clouds are barely resolved with $128^3$, this resolution is sufficient for a box of 40~pc to perform a parameter study. Going towards higher resolution will change the mass fraction in the cold gas, as well as its structure, but it will not affect the qualitative picture. In this paper, we are interested on the global effect of CRs on the two phases of ISM dynamics, and we do not detail the structure of the cold gas quantitatively. We present a resolution study in Appendix \ref{app:resolution} that validates our choice of a $128^3$ resolution for a qualitative study.

We explored the parameter space with the five following parameters: $D_0$ the amplitude of the parallel diffusion coefficient, $k_\mathrm{turb}$ the wave number for turbulence driving, $v_\mathrm{rms}$ the amplitude of the turbulence, $\zeta$ the initial ratio between the CR and the gas pressure, and $\mathcal{M}_\mathrm{A}$ the initial  Alfvénic Mach number.

Table \ref{tab:summary} summarises the different runs adopted for our study (the simulations performed in Appendix \ref{app:resolution} are not reported here). The turbulence injection scale varies from 20~pc for $k_\mathrm{turb}=2$ to 2.5~pc for $k_\mathrm{turb}=16$. The amplitude of the rms velocity  also varies with the driving scale (high driving wave numbers have weaker rms velocity). Most models are initially in a super-Alfv\'enic regime, while the initial turbulent Mach numbers vary from slightly supersonic with $k_\mathrm{turb}=2$ to subsonic with $k_\mathrm{turb}=16$ ($\mathcal{M}\simeq 0.4$). 

If not stated otherwise, we analysed our simulation results at times beyond two turbulent crossing times $t_\mathrm{cross}=L_\mathrm{box}/v_\mathrm{rms}$. We determined the value of the turbulent velocity dispersion $v_\mathrm{rms}$ by running MHD models without CRs and averaging over $\simeq 40$~Myr. Changing the time of analysis or taking a unique absolute time does not change the qualitative results provided that the simulations are analysed once the turbulent motions have settled in a statistically stationary regime.

Similarly to Sect.~\ref{sec:Dcrit}, we can infer a critical diffusion coefficient by assuming that the diffusion timescale is comparable to the eddy turnover timescale, $t_\mathrm{dyn}=L_\mathrm{inj}/v_\mathrm{rms}$, at the injection scale $L_\mathrm{inj}\equiv L_\mathrm{box}/k_\mathrm{turb} $. The critical diffusion coefficient in our setup reads
\begin{equation}
D_\mathrm{crit}= L_\mathrm{inj} v_\mathrm{rms}.
\end{equation}
Since we use a unique value of 40~pc for the size of the computational domain $L_\mathrm{box}$, the critical diffusion coefficient is
\begin{equation}
D_\mathrm{crit}\simeq 1.2\times10^{25} k_\mathrm{turb}^{-1}\left(\frac{v_\mathrm{rms}}{1~\rm{km~s}^{-1}}\right)~\rm{cm}^2~\rm{s}^{-1}.
\label{eq:Dcrit_simu}
\end{equation}
We give in Table~\ref{tab:summary} the values of the critical diffusion coefficient for each setup, which ranges from $1.4\times10^{24}$ to $6.3\times10^{25}$~cm$^2$~s$^{-1}$. We varied $D_0$ from $10^{22}$ to $10^{28}$~cm$^2$~s$^{-1}$ in order to cover the transition regime.

We can also estimate the value of the critical length below which CR are trapped as a function of $D_0$ and $v_\mathrm{rms}$, that is, $L_\mathrm{crit}=D_0/v_\mathrm{rms,}$
\begin{equation}
L_\mathrm{crit}\simeq 3.2\times10^{5}\left(\frac{D_0}{\mathrm{10^{28}~\mathrm{ cm}^{2}~\mathrm{s}^{-1}}}\right)\left(\frac{v_\mathrm{rms}}{1~\rm{km~s}^{-1}}\right)^{-1}~\rm{pc}.
\label{Eq:Lcrit}
\end{equation}
In the case $k_\mathrm{turb}=2$ and $v_\mathrm{rms}=5.7$~km~s$^{-1}$, the critical length varies from about  $5.7$~kpc for $D_0=10^{28}$~cm$^2$~s$^{-1}$ to $5.6\times10^{-3}$~pc for $D_0=10^{22}$~cm$^2$~s$^{-1}$. The condition for CR trapping is thus simply $L_\mathrm{crit}<L_\mathrm{inj}$.

\section{Results\label{section:results}}

\subsection{A fiducial case, $k_\mathrm{turb}=2$}

We first discuss the fiducial set of simulations with a driving scale corresponding to $k_\mathrm{turb}=2$ and an amplitude of the turbulence  $v_\mathrm{rms}\simeq5.7$~km~s$^{-1}$, which corresponds to a turbulent crossing time $t_\mathrm{cross}=6.9$~Myr. We explore five values for the diffusion coefficient: $0-10^{22}-10^{24}-10^{26}-10^{28}$ ~cm$^2$~s$^{-1}$. We also compare the results with a model without CR (labelled No CR). From Eq.~(\ref{Eq:Dcrit}), we expect the critical value for the diffusion coefficient to be $D_\mathrm{crit}\simeq8\times10^{25}\, \rm cm^2 \, s^{-1}$.

Figure~\ref{Fig:time_k2} shows the time evolution of the clumping factor of the  gas density, CR energy density, and Alfv\'enic Mach number for the five above-mentioned simulations with various diffusion coefficients $D_0$ and the model without CRs. The clumping factor of a quantity $X$ is defined as $C(X)=<X^2>/<X>^2$. A clumping factor $C(X)>1$ indicates that the field $X$ has significant variations. The evolution of the clumping factors quickly settles to a quasi-stationary regime within about one turbulent crossing time ($\simeq7$~Myr) once the turbulent motions are established. First, we verified that all simulations show large variations in the density, which indicates that the gas has undergone the thermal instability to settle in the two-phases CNM-WNM medium. The clumping factor of the gas density shows two distinct regimes. In the case where the CR diffusion coefficient is large, $D_0\geq 10^{26}$~cm$^2$~s$^{-1}$, the clumping factor is large ($>4$) and the results are very similar to that of the case without CRs, with an amplitude of a factor $>10^3$ between the minimum and maximum values. On the contrary, with $D_0\leq 10^{24}$~cm$^2$~s$^{-1}$ the clumping factor is close to 1, which indicates that the density field is smoother. The CR energy density shows the opposite behaviour. In the case with $D_0\geq 10^{26}$~cm$^2$~s$^{-1}$, the CR energy density does not vary much, with less than a factor of four to five between the minimum and maximum values, and the clumping factor is very close to unity. For $D_0\leq 10^{24}$~cm$^2$~s$^{-1}$, the CR energy density shows a clumping factor larger than unity corresponding to  large amplitude variations of a factor $>100$ at a given snapshot, which corresponds roughly to the ones found in the density field.   
Last, the Alfv\'enic Mach number variations do not depend on the CR diffusion coefficient, or on the presence of CRs. In addition, we observe that the Alfv\'en speed is independent of the CR pressure, implying that the magnetic fields-density relation is also independent of the CR physics. Since our numerical setup uses the ideal MHD framework, it is natural that the density and the magnetic fields remain well coupled.

The time evolution of the CR energy density shows consistent results, with large variations occurring for $D_0\leq 10^{24}\, \rm cm^2 \, s^{-1}$. Consequently, the gas density variations are tempered by the extra pressure support provided by the gradients in the cosmic-ray pressure, and the maximum gas density is reduced by about one order of magnitude. In the following, we will refer to this regime as the ``trapped CR regime''.

Figure \ref{Fig:map_k2} shows maps of the gas density, CR energy density, and Alfv\'enic Mach number at a time corresponding to $2t_\mathrm{cross}$ for the fiducial runs with  $k_\mathrm{turb}=2$ with CR diffusion. We clearly recover the two different behaviours for $D_0\leq 10^{24}\, \rm cm^2 \, s^{-1}$ and $D_0\geq 10^{26}\, \rm cm^2 \, s^{-1}$ in the density and CR energy density maps. The CR energy and the gas density fields share the same morphology when CRs are trapped, while the density behaves independently of the CR pressure when $D_0\geq 10^{26}\, \rm cm^2 \, s^{-1}$ (CR pressure gradients are shallow). As already mentioned, the Alfv\'enic Mach number shows variations of the same amplitude in all cases.

Figure \ref{Fig:k2_scatter} shows 2D-histograms of the CR energy density, the Alfv\'enic Mach number $\mathcal{M}_\mathrm{A}$, the ratio $\mathcal{M}_\mathrm{s,CR}/\mathcal{M}_\mathrm{s}$, and $\zeta=P_\mathrm{CR}/P$ as a function of the density for all the cells in the computational domain at the same time as in Fig.~\ref{Fig:map_k2}. The colour coding corresponds to the temperature, the magnetic field magnitude, and the sonic Mach number $\mathcal{M}_\mathrm{s}$  (for the two bottom rows), respectively. The correlation between the CR energy density and the gas density is obvious for $D_0\leq 10^{24}\, \rm cm^2 \, s^{-1}$ and the CR fluid behaves as an adiabatic fluid with $P_\mathrm{CR}\propto n^{\gamma_\mathrm{CR}}$, while the CR energy density is insensitive to the density variations for a  larger CR diffusion coefficient. The temperature varies from $\simeq 10^4$~K in the low density regions to $T<100$~K in the high density regions, which indicates that the thermal instability has operated in all models to settle in the expected regime where the cold neutral medium and the warm neutral medium coexist. We note that we neglect CR radiative losses and CR streaming heating, which may affect the thermal balance. These effects will be investigated in future work. As mentioned before, the Alfv\'enic Mach number and magnetic field magnitude do not depend on the value of the CR diffusion coefficient. The ratio $\mathcal{M}_\mathrm{s,CR}/\mathcal{M}_\mathrm{s}$ shows that when CRs are trapped, the CR pressure can substantially modify the effective sound speed of the gas and CR fluid mixture. In the densest regions, the Mach number can be reduced by a factor of two. In the opposite case, the CR pressure has a moderate effect for $D_0\geq 10^{26}\, \rm cm^2 \, s^{-1}$ in the low density regions but no influence at high densities. The behaviour of the pressures ratio $\zeta$ shows also opposite trends. For $D_0\leq 10^{24}\, \rm cm^2 \, s^{-1}$, the CR pressure outmatches the thermal one with increasing density, being the greatest in the highly supersonic regions ($\mathcal{M}_\mathrm{s}>1$). The CNM and WNM are  roughly in pressure equilibrium and then the temperature decreases with increasing density in the CNM, so that the gas effective polytropic index is $\gamma_\mathrm{eff}<1$. The CR energy density is strongly correlated with the density. The CR pressure thus grows faster with the density than the thermal pressure. For $D_0\geq 10^{26}\, \rm cm^2 \, s^{-1}$, $\zeta$ decreases with the density. Since the CR pressure is roughly uniform, $\zeta$ varies with the density as $\zeta \propto 1/n^{\gamma_\mathrm{eff}}$. At high density, the gas is almost isothermal and thus we get $\gamma_\mathrm{eff}\simeq 1$. 

\begin{figure}[t]
        \includegraphics[width=0.5\textwidth]{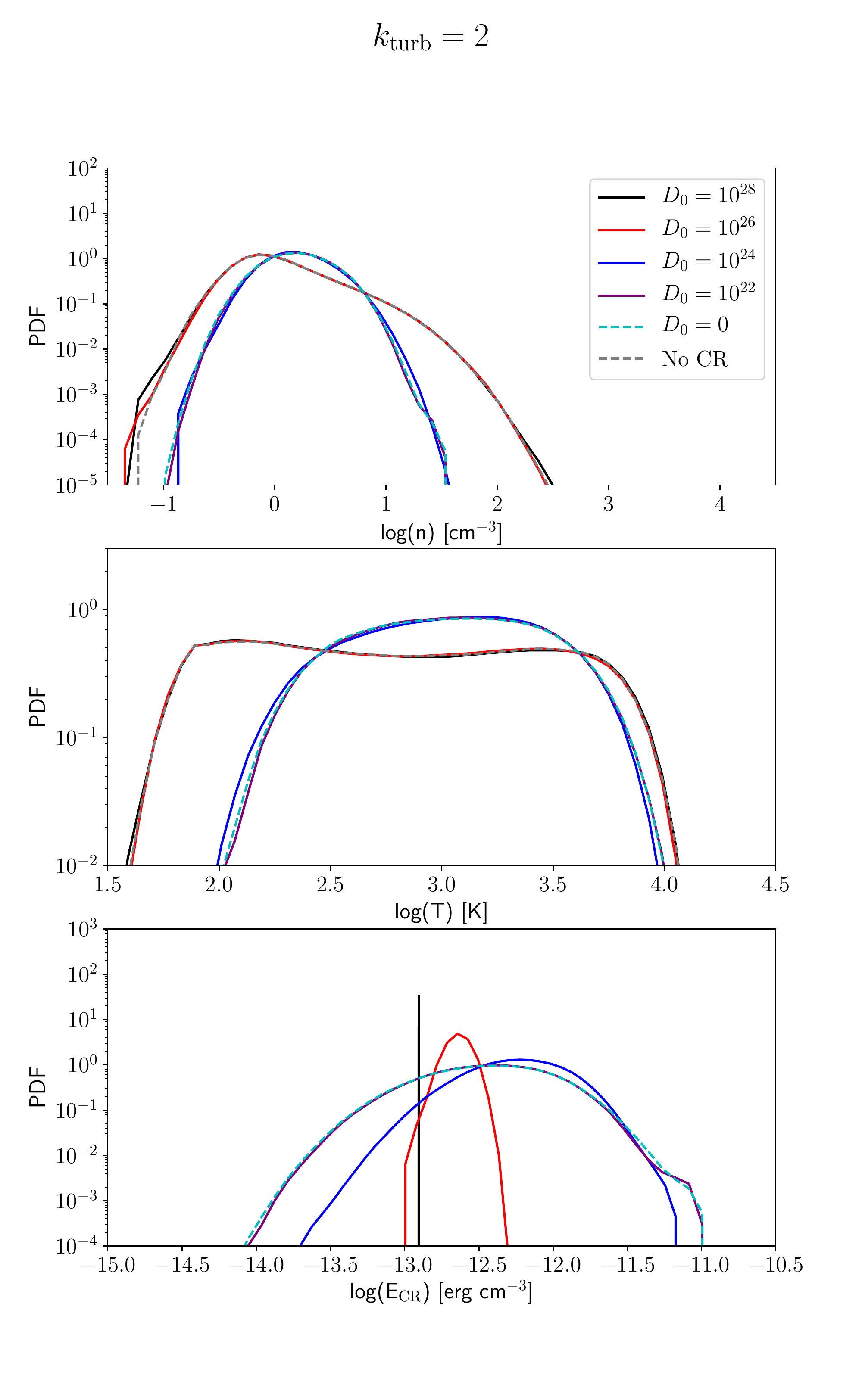}
        \caption{PDF of the gas density (top), temperature (middle), and CR energy density (bottom) for the fiducial runs with $k_\mathrm{turb}=2$ at the same time as in Fig.~\ref{Fig:map_k2}, for  $D_0= 10^{22}$ (purple), $10^{24}$ (blue), $10^{26}$ (red), and $10^{28}$~cm$^2$~s$^{-1}$ (black). Two models with $D_0=0$ (dashed cyan) and without CRs (dashed grey) are also represented for comparison.}
        \label{Fig:pdf_k2}
\end{figure}
\begin{figure}[t]
        \includegraphics[width=0.5\textwidth]{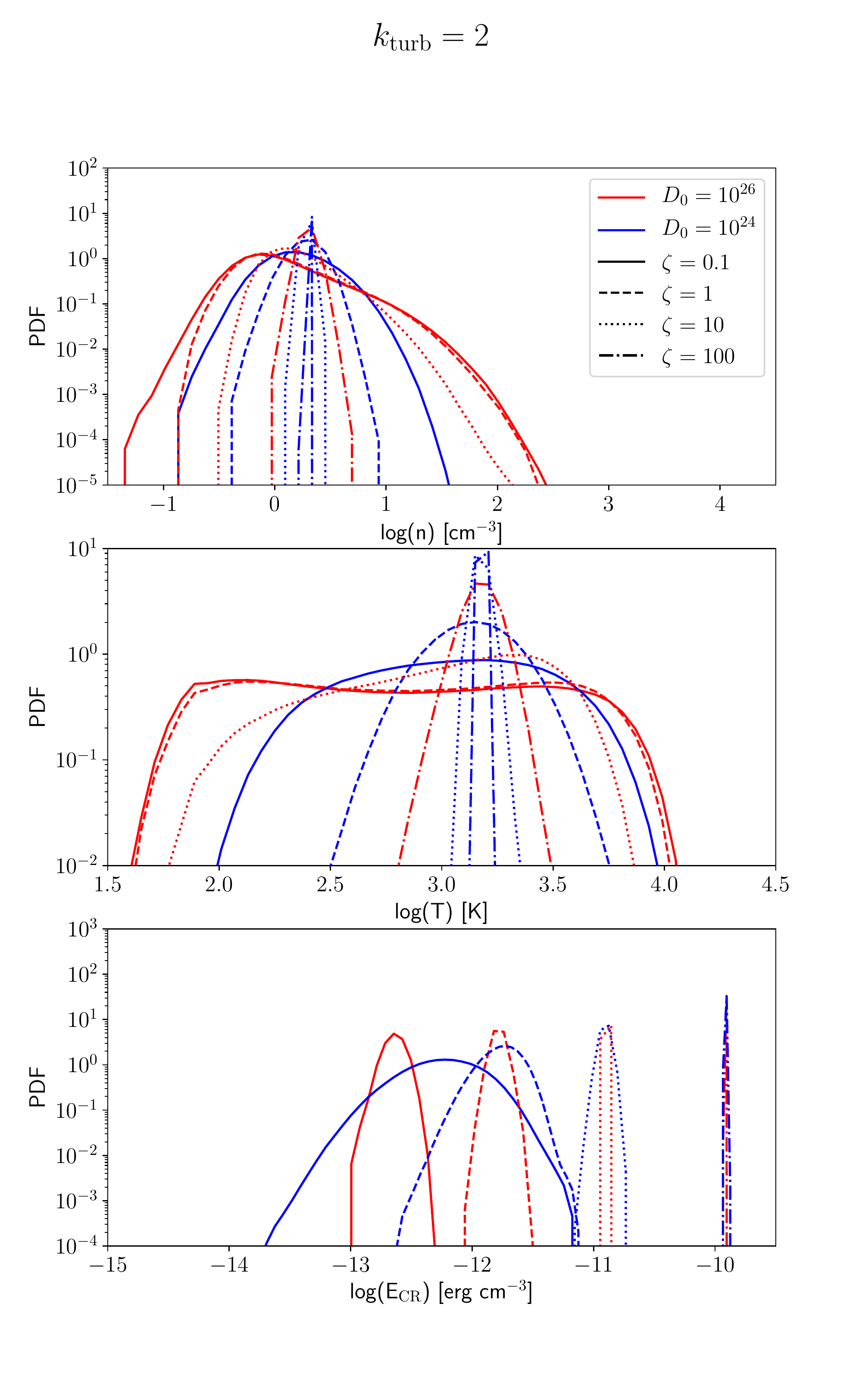}
        \caption{Same as Fig.~\ref{Fig:pdf_k2} for the models with $k_\mathrm{turb}=2$ and $D_0=10^{24}$ (blue), $D_0=10^{26}\, \rm cm^2 \, s^{-1}$ (red). The solid lines represent the fiducial model with $\zeta=0.1$, the dashed line the models with $\zeta=1$, the dotted line $\zeta=10$, and the dashed-dotted line $\zeta=100$.}
        \label{Fig:pdf_b}
\end{figure}
\begin{figure}[t]
        \includegraphics[width=0.5\textwidth]{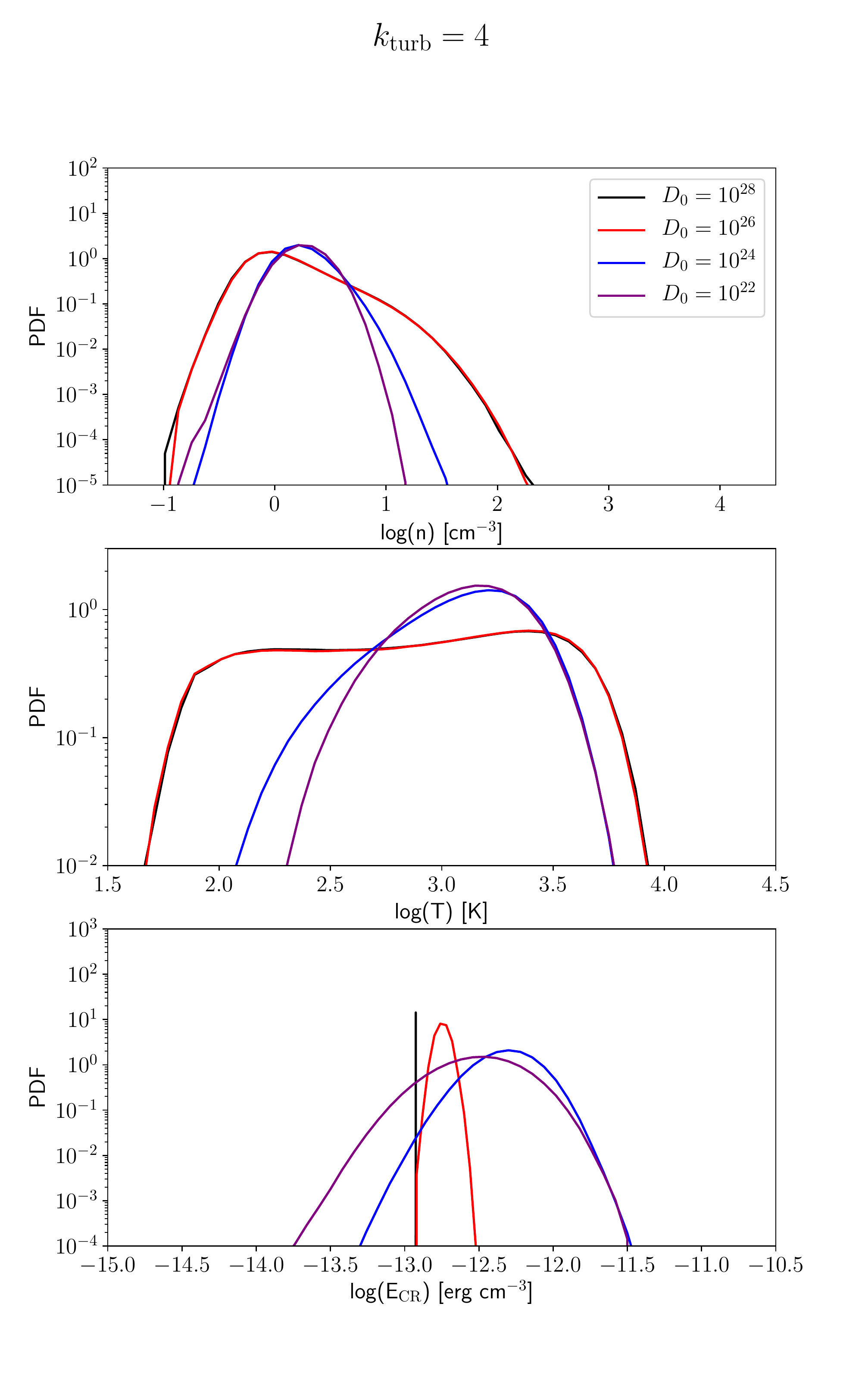}
        \caption{Same as Fig.~\ref{Fig:pdf_k2} for $k_\mathrm{turb}=4$. }
        \label{Fig:pdf_k4}
\end{figure}
\begin{figure}[t]
        \includegraphics[width=0.5\textwidth]{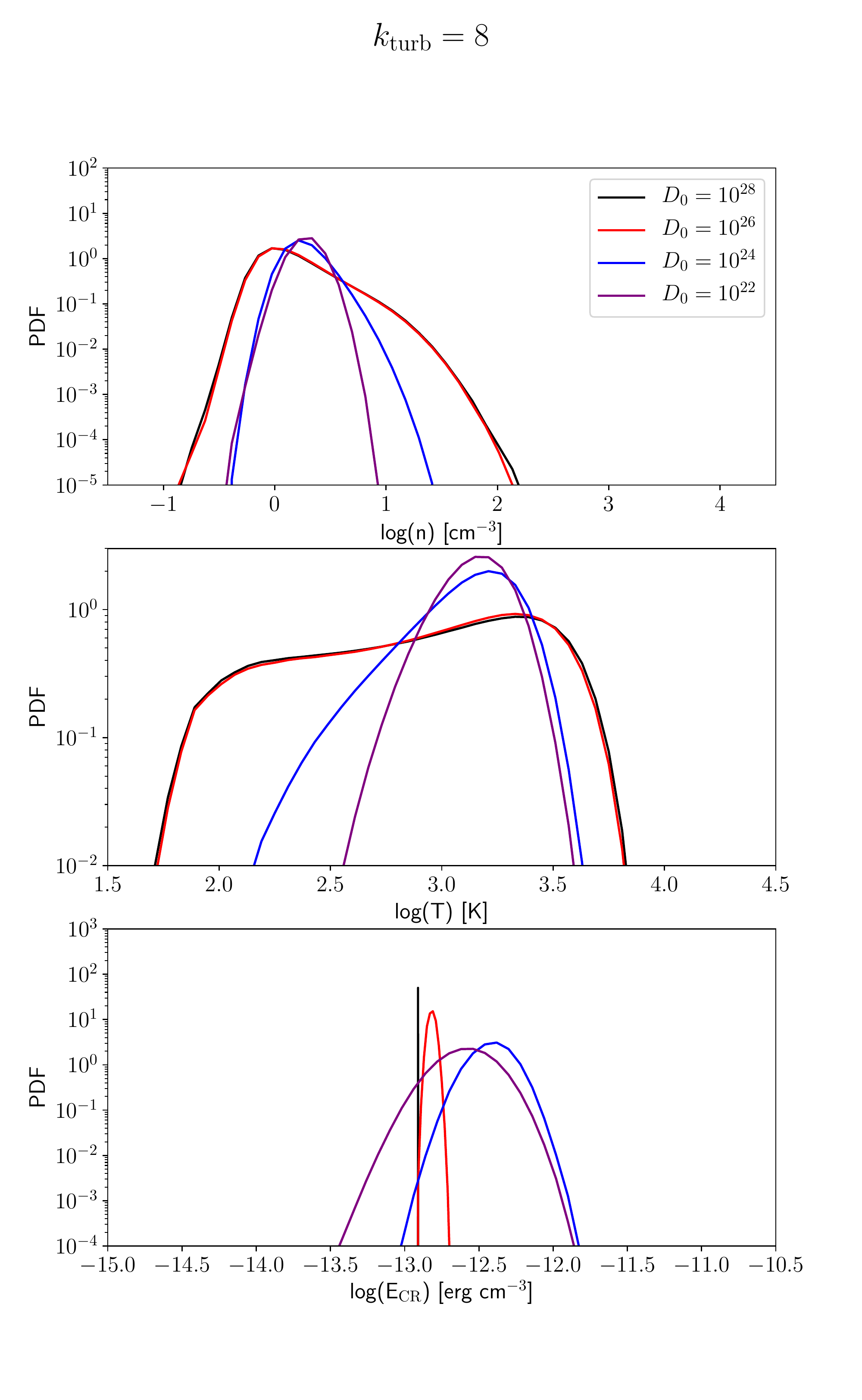}
        \caption{Same as Fig.~\ref{Fig:pdf_k2} for $k_\mathrm{turb}=8$.}
        \label{Fig:pdf_k8}
\end{figure}
\begin{figure}[t]
        \includegraphics[width=0.5\textwidth]{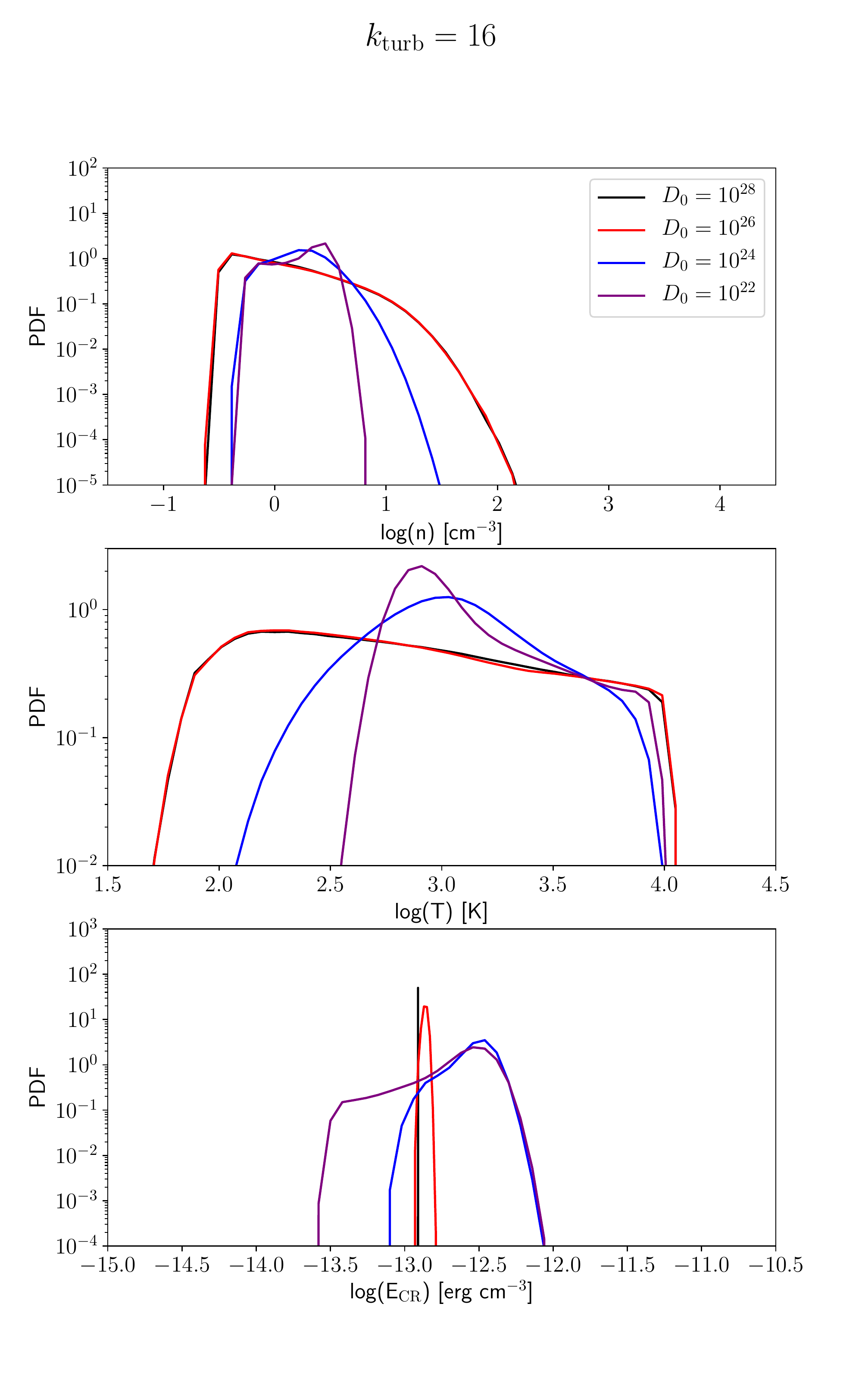}
        \caption{Same as Fig.~\ref{Fig:pdf_k2} for $k_\mathrm{turb}=16$.}
        \label{Fig:pdf_k16}
\end{figure}

The volume-weighted PDF of the gas density, mass-weighted PDF of the gas temperature, and mass-weighted PDF of the CR energy density are shown in Fig.~\ref{Fig:pdf_k2}. In the remainder of the paper, all the PDFs are time-averaged, computed once the calculations have reached $2 t_\mathrm{cross}$. The time average is done over a timescale varying from one to two dynamical times. For comparison, we also show the two extreme runs with $D_0=0$ (no diffusion) and without CRs. There is again an obvious distinction depending on whether the CR diffusion coefficient is higher or lower than the critical value. The gas density and temperature PDFs are wider in models with $D_0\geq 10^{26}\, \rm cm^2 \, s^{-1}$, suggesting that the gas is more efficiently compressed (or expanded) compared to the lower diffusion coefficient values. From first principles, one would have expected that the gas gets more compressed since the mixture of gas and CRs has a global adiabatic index $4/3<\gamma<5/3$ (see Eq.~\ref{Eq:cs}). Nevertheless, the gas compression is primarily dependent on the thermal budget resulting from the heating and cooling processes included. As we just discussed, the effective polytropic index of the gas is even lower than $4/3$ \citep[e.g.][]{wolfire:03}. In the trapped CR regime, the CR pressure dominates so that the effective polytropic index of the mixture is close to the CR one and prevents strong compression with respect to the case where thermal pressure  dominates ($D_0\geq 10^{26}\, \rm cm^2 \, s^{-1}$).
The temperature variations are reduced in the models with $D_0\leq 10^{24}\, \rm cm^2 \, s^{-1}$. In the case without CRs, the temperature PDF barely shows the two expected peaks for the CNM and the WNM  mixture at $T\simeq60$~K (CNM phase) and $T\simeq5000$~K (WNM phase). The maximum temperature increases by adiabatic heating to reach a higher value than the initial one, while the various thermal processes included in the heating and cooling  function lead to an efficient cooling in the high density regions.  
The CR energy density PDF is more sensitive to the diffusion coefficient than the gas density and temperature ones. The CR energy variation amplitudes are anti-correlated with those of the gas density and temperature. For $D_0= 10^{28}\, \rm cm^2 \, s^{-1}$, the CR energy shows no variation and remains uniform at an energy density corresponding to the initial state. The CR and gas fluids are decoupled and CRs propagate freely in the gas. The CR energy density begins to show variations of less than a decade for $D_0=10^{26}\, \rm cm^2 \, s^{-1}$. CR pressure gradients thus develop, but have no effect on the gas flow as indicated in the density PDF. For $D_0=10^{24}\, \rm cm^2 \, s^{-1}$, the CR energy density PDF is much wider ($\simeq2.5$ decades at PDF$=10^{-3}$), so that CR pressure gradients develop. The peak of the PDF shifts towards higher values compared to the initial value. The trapped CRs are compressed and then diffuse slowly so that the peak moves towards higher CR energy densities. Interestingly, the CR energy density PDFs vary substantially between $D_0=10^{22}\, \rm cm^2 \, s^{-1}$ and $D_0=10^{24}\, \rm cm^2 \, s^{-1}$ while this is not the case for the gas density and the temperature. For  $D_0=10^{22}\, \rm cm^2 \, s^{-1}$, the three PDFs match closely the ones of the model without CR diffusion, that is where the CRs are perfectly trapped with the fluid. In the opposite case, the gas density and temperature of the case where $D_0= 10^{28}\, \rm cm^2 \, s^{-1}$ correspond to those of the run without CRs. 

The picture we can draw from these fiducial runs is the following: for diffusion coefficients $D_0\leq 10^{24}\, \rm cm^2 \, s^{-1}$, CRs are trapped and can provide extra pressure support to counterbalance the increase in gas density due to adiabatic compression and/or shocks, and the cooling becomes less efficient. In the model with $D_0\leq 10^{24}\, \rm cm^2 \, s^{-1}$, the thermal instability is thus less efficient in segregating the gas in the mixture of WNM and CNM. As already mentioned, we neglect in this work CR streaming and radiative losses. Further work will investigate the impact of these effects, in particular in the sub-Alfv\'enic regions, on the qualitative picture we present here.
We observe that the width of the gas density PDF increases with the CR diffusion coefficient. As mentioned in the Introduction, the width of the gas density PDF can be related to the fluid sonic Mach number. In our model, the fluid is made of a mixture of gas and CRs and the extra CR pressure support increases the effective sound speed of the mixture. As a consequence, the effective Mach number decreases as shown in Fig.~\ref{Fig:k2_scatter} and the PDF gets tighter.

We finally note that given our choice of parameters, the critical length for CR trapping was $\simeq57$~pc (larger than the box size of 40 pc) for $D_0= 10^{26}\, \rm cm^2 \, s^{-1}$, which is indeed verified by our numerical results since the results show no evidence of CR trapping with this diffusion coefficient. 

\subsection{Parameter study}
In this section, we explore the parameter space by varying the CR initial pressure, the driving scale, the driving amplitude, and the initial magnetisation level. 

\subsubsection{Effect of the CR initial pressure}
The first parameter we test is the initial ratio $\zeta$ between the CR pressure and the gas pressure. In the fiducial case, we used $\zeta=0.1$. We now explore three additional cases, with $\zeta=1$, 10, and 100. We used the same initial conditions and turbulence driving properties as in the fiducial runs (only the CR pressure increases). For each value of $\zeta$, we performed two simulations with $D_0=10^{24}$ and $D_0=10^{26}\, \rm cm^2 \, s^{-1}$, which surround the critical value $D_\mathrm{crit}$.

Figure \ref{Fig:pdf_b} shows the PDFs of the gas density, the temperature, and the CR energy density for the fiducial cases and the six runs with $\zeta=1$, 10, and 100. For a diffusion coefficient $D_0=10^{26}\, \rm cm^2 \, s^{-1}$ and $\zeta\leq1$, the gas density and temperature PDFs are similar in width. We note some differences in the maximum density (and thus minimum temperature), which is lower (respectively, higher) with $\zeta=1$. The CR energy density is shifted by about one decade towards a higher value with $\zeta=1$, which corresponds to the initial difference between the CR pressure. The width of the PDF is narrower with $\zeta=1$, which indicates that the CRs are less sensitive to the gas flow fluctuations if their pressure is larger. Nevertheless, the CR pressure has a slight effect on the gas flow as indicated by the maximum density, which is a little bit lower. The CR pressure is strong enough to provide an extra support. For $\zeta>1$ and  $D_0=10^{26}\, \rm cm^2 \, s^{-1}$, the three PDFs become tighter as $\zeta$ increases, up to the point where the CR pressure prevents any compression of the gas and thus the development of the thermal instability. As the gas density PDF can be well approximated by a log-normal distribution, which width is determined by the sonic Mach number, increasing the CR pressure leads to a decrease of the modified sound speed $c_\mathrm{s,CR}$ and, hence, to a decrease in the width of the PDF. In the $D_0=10^{24}\, \rm cm^2 \, s^{-1}$ case, the difference between $\zeta=0.1$ and $\zeta\geq1$ is striking. The larger CR pressure inhibits the development of the thermal instability and the gas remains in a state that is thermally unstable. CNM gas formation, where $n>10$~cm$^{-3}$ , is strongly altered if CRs are in pressure equilibrium with the gas (roughly energy equipartition) and if they get trapped. In this case, the transition between the two regimes is much more abrupt. If the initial CR pressure exceeds the thermal one, the gas does not evolve much and the thermal instability does not develop even though it is initially in a state that is thermally unstable. The classical criterion for the thermal instability can roughly be can be summarised as $(dP/d\rho)<0$ \citep{field:1965}. Our results indicate that this criterion must be revised if one accounts for the CR pressure. Values of $\zeta \ge 1$ are to be expected in the ISM surrounding CR sources
while CRs are freshly injected in. Our results show that CR sources can strongly modify the ambient ISM dynamics even if the CR diffusion coefficient is larger than $D_{\rm crit}$. This aspect has to be accounted for in simulations including supernova (SN) feedback.

\subsubsection{Effect of the driving scale}

\begin{figure}[t]
        \includegraphics[width=0.5\textwidth]{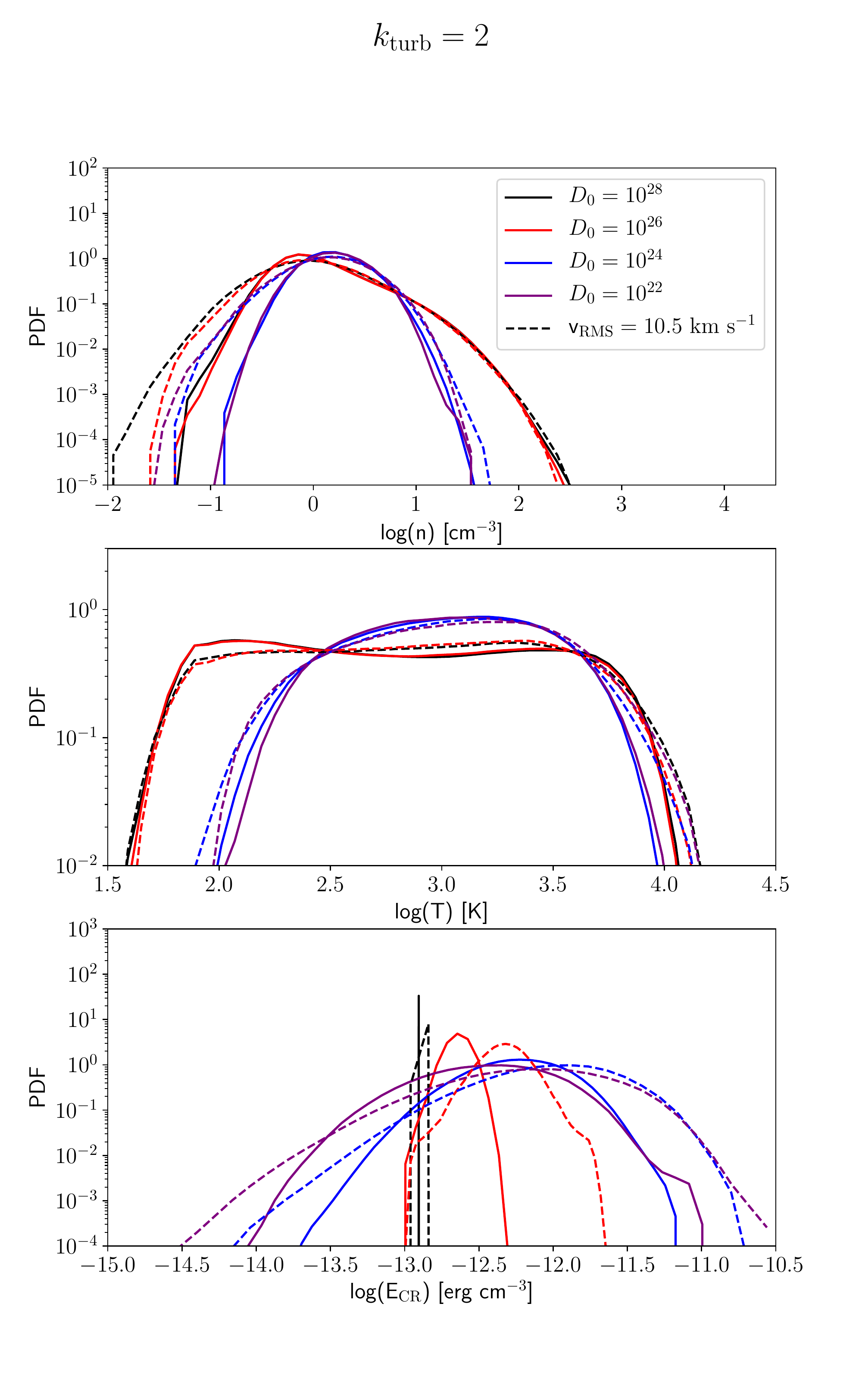}
        \caption{Same as Fig.~\ref{Fig:pdf_k2} for $k_\mathrm{turb}=2$, but different amplitude in the turbulence driving. The solid lines represent the fiducial case results with $v_\mathrm{rms}=5.7$~km~s$^{-1}$  and the dashed lines the ones obtained with  $v_\mathrm{rms}=10.5$~km~s$^{-1}$.}
        \label{Fig:pdf_k2_rms}
\end{figure}

We present in this section the results obtained by running models with the same initial conditions as in the fiducial case, but where we vary the scale for turbulence driving. We explore three smaller driving scales, corresponding to $k_\mathrm{turb}=4$, 8, and 16. The aim is to probe the effect of a smaller injection scale for the turbulence. We keep the driving amplitude of the acceleration constant between the different models so that the rms Mach number decreases with increasing $k_\mathrm{turb}$.  We thus indirectly test at the same time the effect of varying the turbulence amplitude. 

From the measured value of $v_\mathrm{rms}$ reported in Table~\ref{tab:summary}, the critical length for CR trapping is 8.2, 11, and 17.8~pc for $k_\mathrm{turb}=4,8,16$ respectively (and for $D_0=10^{24}\, \rm cm^2 \, s^{-1}$), while the turbulence injection scale varies from 10~pc to 2.5~pc. This means that for the $k_\mathrm{turb}=8$ and 16 models, the driving scale $L_\mathrm{inj}$ is less than the expected critical length $L_\mathrm{crit}$ for $D_0=10^{24}\, \rm cm^2 \, s^{-1}$. In addition, since the turbulent energy decreases with increasing $k_\mathrm{turb}$, $D_0=10^{24}\, \rm cm^2 \, s^{-1}$ is closer to the critical diffusion coefficient reported in Table \ref{tab:summary} than in the fiducial runs.

Figures \ref{Fig:pdf_k4}, \ref{Fig:pdf_k8}, and \ref{Fig:pdf_k16} show the gas density, temperature, and CR energy density PDFs for the models with varying driving scale $k_\mathrm{turb}$ (4, 8, and 16, respectively). In all cases, we recover the two regimes, but the transition shifts towards lower values of $D_0$ as $k_\mathrm{turb}$ increases, as expected in the qualitative analysis above. For $D_0\geq 10^{26}\, \rm cm^2 \, s^{-1}$, the gas always settles in the two phases mixture of CNM and WNM. The width and global shape of the gas density and temperature remain similar for $k_\mathrm{turb}=4$ and 8, though with gas that is denser and warmer in the WNM as $k_\mathrm{turb}$ increases (compression is simply less efficient). In the $k_\mathrm{turb}=16$ case, the density PDF is shallower at high density. Most of the gas remains at a low density because the turbulence is not strong enough to push the gas along the Hugoniot curve towards a higher density. Nevertheless, we observe that the PDF of the temperature does not reflect that of the density in the $k_\mathrm{turb}=16$ models, with an excess of low temperature gas. The spread in density at a given temperature is indeed larger in the cases where the turbulence is stronger as the density fluctuations may get destroyed faster, so that the thermal instability cannot develop quickly and cooling becomes less efficient. In the case of low turbulence, the gas remains predominantly in thermal equilibrium and the fraction of cold gas increases \citep[e.g.][]{audit:05,seifried:11,gazol:13}. Overall, the PDF of the CR energy density is more sensitive to the driving scale. For $D_0\geq 10^{26}\, \rm cm^2 \, s^{-1}$, the width gets narrower as the turbulence level decreases. In the trapped CR regime, the width of the CR energy density PDF scales with the gas density and thus gets also narrower at higher $k_\mathrm{turb}$. Lastly, we do not observe any particular effect when the driving scale is less than the critical length for $D_0=10^{24}\, \rm cm^2 \, s^{-1}$. The transition to the trapped CR regime remains on the same parameter range.

\subsubsection{Effect of the driving amplitude or sonic Mach number}

The next experiment  consists in increasing by a factor $\simeq 2$ the amplitude of the turbulence driving in the fiducial case ($k_\mathrm{turb}=2$, i.e. the driving scale remains the same contrary to the preceding section) to reach $v_\mathrm{rms}=10.5$~km~s$^{-1}$.  We performed four simulations with a CR diffusion coefficient ranging from $D_0=10^{22}$ to $D_0=10^{28}$~cm$^2$~s$^{-1}$.

Figure \ref{Fig:pdf_k2_rms} shows the gas density, temperature, and CR energy density PDFs with $v_\mathrm{rms}=10.5$~km~s$^{-1}$. The PDFs of the fiducial case with $v_\mathrm{rms}=5.7$~km~s$^{-1}$ are also shown for comparison. From the gas density PDF, the main effect of a stronger turbulence is to produce lower density regions for all values of $D_0$, as a result of larger expansion. The high gas densities are not very sensitive to the amplitude of the turbulence driving. As a consequence, the temperature is also not very sensitive to the amplitude of the turbulence. Finally, the CR energy density shows the same qualitative behaviour as in the fiducial case, with trapped CRs for  $D_0\leq 10^{24}\, \rm cm^2 \, s^{-1}$. The PDFs are nevertheless different, with larger width and higher peak values with stronger turbulence. The peak is shifted toward higher energy density as a result of stronger compression by turbulent motions. The minimum value of the CR energy density decreases with increasing $v_\mathrm{rms}$ as a result of local expansions, which create low energy density that cannot be refilled  with surrounding CRs since diffusion is less efficient. 

\begin{figure}[t]
        \includegraphics[width=0.5\textwidth]{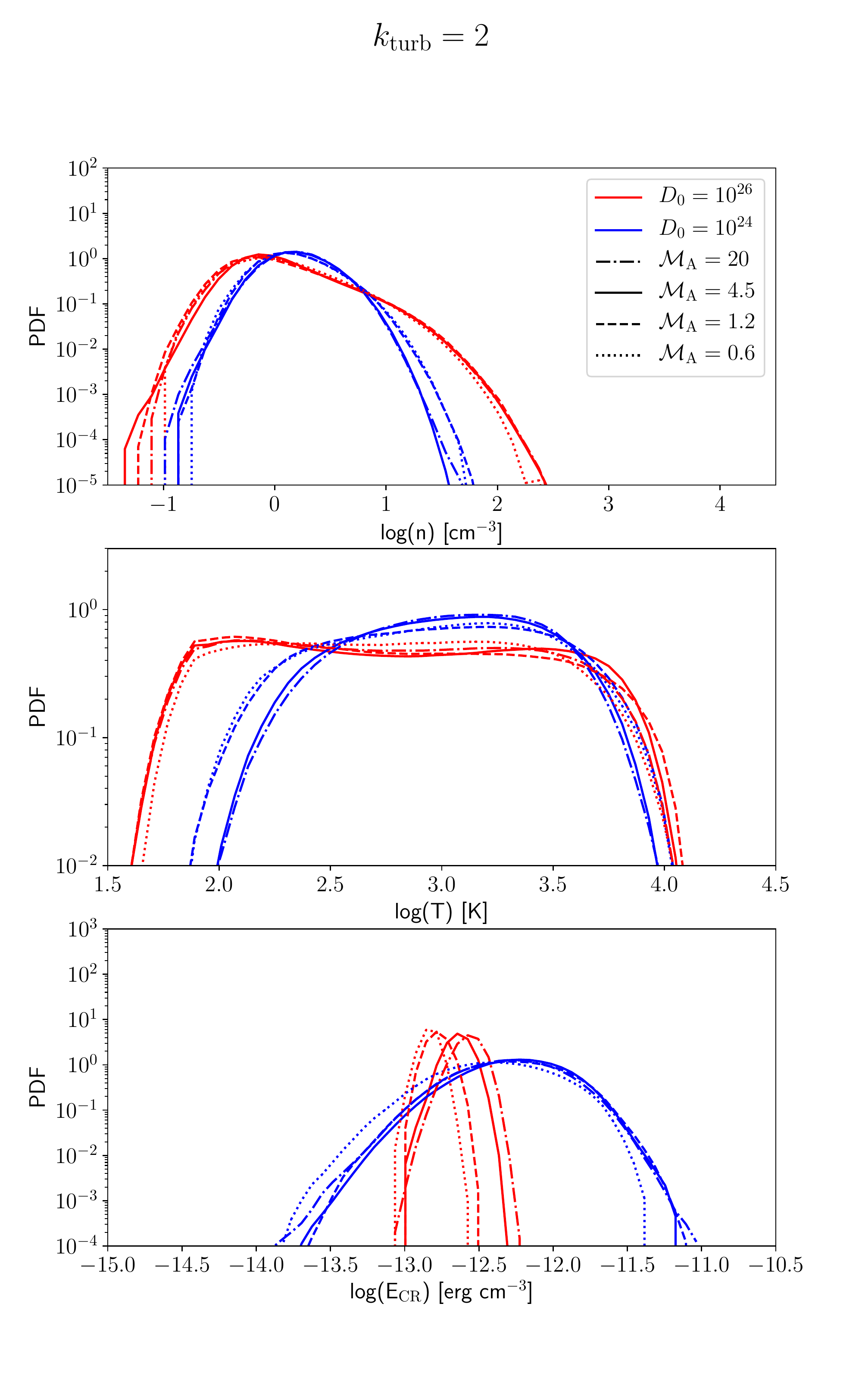}
        \caption{Same as Fig.~\ref{Fig:pdf_k2} for $k_\mathrm{turb}=2$, but different amplitude of the initial magnetic field amplitude. The solid lines represent the fiducial case  with $\mathcal{M}_\mathrm{A}=4.5$,  the dashed lines $\mathcal{M}_\mathrm{A}=1.2$, the dotted lines $\mathcal{M}_\mathrm{A}=0.6$, and the dashed-dotted lines $\mathcal{M}_\mathrm{A}=20$. The blue colour represents results with $D_0=10^{24}$~cm$^2$~s$^{-1}$, the red $D_0=10^{26}$~cm$^2$~s$^{-1}$.}
        \label{Fig:pdf_k2_Ma}
\end{figure}

\subsubsection{Effect of the magnetic field amplitude or Alfv\'enic Mach number}
In this last experiment, the dependency on the initial Alfv\'enic Mach number $\mathcal{M}_\mathrm{A}$ is explored by changing the value of the initial magnetic field amplitude. We compared the fiducial case results, with $\mathcal{M}_\mathrm{A}=4.5$ (super-Alfv\'enic), with the ones obtained with $\mathcal{M}_\mathrm{A}=20$ (super-Alfv\'enic, weak magnetic fields),  $\mathcal{M}_\mathrm{A}=1.2$ (trans-Alfv\'enic), and $\mathcal{M}_\mathrm{A}=0.6$ (sub-Alfv\'enic). We recall that magnetic field amplitude increases with decreasing $\mathcal{M}_\mathrm{A}$. We use two values for the CR diffusion coefficient, $D_0=10^{24}$~and $D_0=10^{26}$~cm$^2$~s$^{-1}$

Figure \ref{Fig:pdf_k2_Ma} shows the PDFs of the eight different runs.  Again, the qualitative behaviour is not modified, with a transition to the regime of trapped CRs for $D_0\leq 10^{26}$~cm$^2$~s$^{-1}$. The gas density and temperature PDFs do not depend much on $\mathcal{M}_\mathrm{A}$ for $D_0= 10^{26}$~cm$^2$~s$^{-1}$. A noticeable feature is the width of the gas density PDF that is the narrowest with $\mathcal{M}_\mathrm{A}=0.6$. This is the expected behaviour  since stronger magnetic fields provide support that softens the turbulent motions \citep[e.g.][]{molina:12}. The CR energy density PDF gets wider as $\mathcal{M}_\mathrm{A}$ increases. The magnetic field lines are indeed more tangled in the super-Alfv\'enic case, which reduces the effective diffusion coefficient. For $D_0=10^{24}$~cm$^2$~s$^{-1}$, we see clearly two different regimes as a function of $\mathcal{M}_\mathrm{A}$, and the results are qualitatively different from the case with $D_0= 10^{26}$~cm$^2$~s$^{-1}$. In the super-Alfv\'enic regime, the gas density and temperature PDF are narrower, meaning that as the magnetisation increases the gas gets more compressed. As we just mentioned, the effective CR diffusion coefficient decreases in the super-Alfv\'enic regime. CRs are thus more efficiently trapped and CR pressure gradients get stronger. As a consequence, the CR pressure support increases and counterbalances the effect of the increase in magnetisation level. Clearly, the coupled effect of the magnetisation level and of CR propagation is non-linear and should be investigated in future works with more details partly based on some specific CR transport modelling (see Sect.~\ref{S:LST}).

\section{Discussion\label{section:discussion}}
 
\subsection{Dependence on the parameters}

We have seen in the previous section that the critical diffusion coefficient for CR trapping  $D_{\rm crit}$ is of the order of $10^{24}-10^{25}$~cm$^2$~s$^{-1}$ and is in good agreement with the theoretical expectations. 
In addition, we have observed a strong effect of the initial pressure ratio $\zeta$. For large initial CR pressure, $\zeta = 10$, the gas and CR mixture behaves differently even if CRs are not trapped. The non-thermal support provided by the CR pressure may be very efficient at preventing the gas compression by turbulent motions so that the thermal instability is inhibited in CR-pressure-dominated flow. We show in the following discussion that  CR sources and  CR acceleration at small scales are possible mechanisms for local enhancements of the CR pressure, which might then change the gas dynamics in the environment of these sources.

 In addition, the initial Alfv\'enic Mach number has an effect on the gas compression. In the case of initial super-Alfv\'enic Mach numbers, the magnetic field lines are tangled and the CR diffusion tends to be isotropic with a reduced effective coefficient. The maximum CR energy density is thus lower and the CRs' pressure gradients are weaker than in the case of sub-Alfv\'enic turbulence. We note that we retrieve this behaviour when an isotropic diffusion coefficient is used (see Appendix \ref{appendix:isotropic}). We have seen in Sect.~\ref{section:analysis} and Fig. \ref{Fig:Ma_MC} that the ISM MHD turbulence can be either super- or sub-Alfv\'enic, and we expect to recover both the isotropic and anisotropic diffusion regimes. We also discuss in Sect. \ref{S:LST} some models for the diffusion coefficient depending on the large-scale Alfv\'enic Mach number. 

\subsection{Possible mechanisms for CR diffusion coefficient variation}
At a first glance it might be hopeless to investigate ISM dynamics under the effect of CRs because the usual ISM diffusion coefficient is $D_{\rm ISM} \gg D_{\rm crit}$.
However, several physical processes can contribute to decrease the effective CR diffusion coefficient below $D_{\rm ISM}$ and possibly below $D_{\rm crit}$ in some specific ISM environments. We provide a qualitative discussion of these processes below but as this first study aims only at a parametric investigation of the effect of CR pressure over ISM dynamics, we postpone to an associated paper (paper II) a detailed study of these specific diffusion regimes. We also caution the reader that some processes we invoke in the following may be at scales we do not consider in our previous numerical experiments.

\subsubsection{Large-scale turbulence models}\label{S:LST}
Large-scale motions in the ISM are sources of turbulence \cite[][]{maclow:04}. Among the potential processes that are sources of turbulence we find magneto-rotational instabilities produced by differential galactic rotation motions, supernova explosions either isolated or through a cumulative effect in superbubbles, and gravitational instabilities induced by galactic arm motions. At smaller galactic scales other sources of turbulence injection can contribute, especially protostellar jets, stellar wind feedback, or ionisation front motions associated to H$_{\rm II}$ regions.  All these processes force turbulent motions with different geometries, either solenoidal or compressible or a combination of both, at some preferred scale $L_{\rm inj}$. Once fluid motions are forced at $L_{\rm inj}$ , energy cascades down to smaller scales. This cascade can be described by different models. As the ISM is magnetised, large-scale motions can be described approximately using a MHD turbulent model. At the scales under investigation, we have seen in Sect.~\ref{section:turbulent} that turbulent motions are transonic in the WNM and supersonic in the CNM, so compressible. MHD  compressible turbulence has been mainly studied by means of numerical simulations \citep{lithwick:01, cho:02}. \\

In particular, \citet{cho:02} isolate different regimes of MHD turbulence depending on the regime of Alfv\'enic Mach number. Based on this turbulence model, \citet{yan:08} propose a derivation of CR diffusion coefficients induced by these large-scale-injected turbulent motions. Let us recall their main results.

\paragraph{Super-Alfv\'enic regime.} If ${\cal M}_{\rm A} > 1$, then large-scale motions are hydrodynamical and follow a Kolmogorov law. At a scale $\ell_{\rm A} = L_{\rm inj}/{\cal M}_{\rm A}^3$ the turbulence becomes trans-Alfv\'enic. At scales below $\ell_{\rm A}$ the turbulence is MHD and can be decomposed into three MHD mode cascades: Alfv\'en and slow magnetosonic cascades are anisotropic and follow a Goldreich-Sridhar model \citep{goldreich:95}, while the fast magnetosonic cascade is isotropic and follows a Kraichnan model \citep{cho:02}. CR diffusion in super-Alfv\'enic turbulence is isotropic, hence $D=D_\parallel = D_\perp$ but depends on $\lambda_\parallel$, the CR mean free path parallel to the mean magnetic field. If $\lambda_\parallel > L_{\rm inj}$, the effective diffusion coefficient is $D = \ell_{\rm A} v/3$ while if $\lambda_{\parallel} < L_{\rm inj}$, $D= \lambda_{\parallel} v/3$, where $v =\beta_\mathrm{c} c$ is the CR velocity. The calculation of $\lambda_\parallel$ in partially ionised phases is complex and involves a calculation of the turbulence damping scale by ion-neutral collisions. \citet{xu:16} find a characteristic U shape for $\lambda_\parallel$ with respect to the CR energy. It appears that in the WNM and CNM phases, GeV CRs have Larmor radii smaller than MHD turbulence damping scales and are in the regime $\lambda_\parallel \gg L_{\rm inj}$. Alfv\'enic Mach numbers in these partially ionised phases are not well known but it is possible to define a critical Alfv\'{e}nic Mach number using Eq.~(\ref{Eq:Dcrit}),

\begin{equation}
{\cal M}_{\rm A, crit} \equiv \left({v L_{\rm inj} \over 3 D_{\rm crit}}\right)^{1/3} \simeq 46~\left(\frac{\beta_{\rm c} L_{\rm inj}}{1~{\rm pc}}\right)^{1/3}\left(\frac{L}{1~{\rm pc}}\right)^{-{(q+1)\over3}}. 
\end{equation}

If $\cal{M}_{\rm A} \ge \mathcal{M}_\mathrm{A,crit}$, CRs can have dynamical effects. Using $L_{\rm inj} = L/2=20$ pc, q=1/3 or 1/2 as in our fiducial simulation we find for 1 GeV CRs ${\cal M}_{\rm A, crit} \simeq 23$ for q=1/3 and ${\cal M}_{\rm A, crit} \simeq 19$ for q=1/2. From Fig.~\ref{Fig:Ma_MC} we can deduce that this regime is marginally obtained for the gas density and scale length considered in this work. However, smaller critical Alfv\'enic Mach numbers are expected if the turbulence is injected at smaller scales (at higher $k_{\rm turb}$).

\paragraph{Sub-Alfv\'enic regime.} If ${\cal M}_{\rm A} < 1$, magnetic fields dominate gas dynamics at the injection scale. The turbulence is weak and cascades in the direction perpendicular to the mean magnetic field \citep{galtier:00}. Below a scale $\ell_{\rm w} = L_{\rm inj} {\cal M}_{\rm a}^2$ the Goldreich-Sridhar scaling prevails again. Sub-Alfv\'enic CR diffusion is no longer isotropic. The perpendicular diffusion coefficient depends again on $\lambda_\parallel$ \citep{yan:08}. If $\lambda_\parallel > L_{\rm inj}$ then $D_\perp = D_\parallel {\cal M}_{\rm A}^4 L_{\rm inj}/\lambda_\parallel$ , while if $\lambda_\parallel < L_{\rm inj}$, 
$D_\perp =D_\parallel {\cal M}_{\rm A}^4$; in each case, $D_\perp \ll D_\parallel$. The parallel diffusion coefficient is $D_\parallel = \lambda_{\parallel} v/3$. In that case one can define a critical parallel mean free path $\lambda_{\parallel, \rm{crit}}$
\begin{equation}
\lambda_{\parallel, \rm{crit}} \simeq 3.1\times10^{13}~\rm{cm}~\beta_\mathrm{c}^{-1} \left(\frac{L}{1~{\rm pc}}\right)^{q+1} ,
\end{equation}
 which in our fiducial case using Eq.~(\ref{Eq:Dcrit}) is $\lambda_{\parallel, \rm{crit}} \simeq 4.9\times 10^{15}$ cm for q=1/3 or $\lambda_{\parallel, \rm{crit}} \simeq 9.0 \times 10^{15}$ cm for q=1/2. Calculations by \citet{xu:16} indicate that at GeV energies $\lambda_\parallel > L_{\rm inj}$ so likely larger than $\lambda_{\parallel, \rm{crit}}$. We do not expect any strong CR effect in this regime. \\

\subsubsection{CR streaming and turbulence generation}

Low energy CRs are injected at the end of a supernova remnant's (SNR) lifetime likely when the forward shock becomes trans-sonic but possibly before when the SNR enters  the radiative phase when shock acceleration stops. As most SNRs explode behind spiral arms and usually in groups (thus forming superbubbles), they produce a mean CR density (or pressure) gradient in the galaxy both in vertical and radial directions. CRs while propagating at the speed of light along the mean magnetic field can trigger MHD perturbations preferentially along the magnetic field lines through the so-called streaming instability \citep{skilling:71}. Perturbations generated by CRs are resonant and have typical wave numbers $k_{\rm res}\sim r_{\rm g}^{-1}$, where $r_{\rm g} =E/eB$ is the CR gyroradius for a CR kinetic energy $E$ and a charge $e$ in a magnetic field of strength $B$. The streaming instability growth rate $\Gamma_{\rm g}$ is controlled by the CR density or pressure gradient along the field lines. In the ISM, we can expect typical CR pressure gradients of the order of $P_{\rm CR}' \sim P_{\rm CR}/L_{\rm CR}$, where $L_{\rm CR}$ is the CR gradient scale. A reference value is $P_{\rm CR, ref}' \simeq 5.2\times 10^{-33}~{\rm dyne~cm^{-3}}~P_{\rm CR, eVcc}/L_{\rm CR, 100 pc}$, where $P_{\rm CR, eVcc}$ is the CR pressure in units of $1~\rm{eV~cm^{-3}}$ and $L_{\rm CR}$ is the CR gradient scale in units of 100 pc. In the Milky Way, one expects $P_{\rm CR} ' \le P_{\rm CR, ref}'$ as the typical CR variation scale lengths are larger than the galactic disc height.  \\

CRs can inject magnetic perturbations at small scale lengths through the streaming instability. The level of the perturbations saturate due to a balance of the growth rate with a damping rate. Depending on the ISM phase under consideration the dominant damping process changes \citep{wiener:13, nava:16}. In partially ionised phases, the dominant damping process is due to collisions between ions and neutrals. At GeV energies, the CR self-generated waves are at too high frequencies to prevent ions from being decoupled from neutrals. The damping rate is $\Gamma_{\rm in} = \nu_{\rm in}/2$, where again $\nu_{\rm in}$ is the ion-neutral collision frequency. We take $\nu_{\rm in} \simeq 1.4 ~10^{-9} n_{\rm n}~ (T/100 \rm{K})^{0.5}~\rm{cm^3~s^{-1}}$ and $\nu_{\rm in} \simeq 1.6 ~10^{-9} n_{\rm n}~\rm{cm^3~s^{-1}}$ with $n_{\rm n}$ the number density of neutrals, in the WNM and the CNM, respectively \citep{jean:09}. In the decoupled ion-neutral regime the streaming instability growth rate (in $\rm{s^{-1}}$) can be written as $\Gamma_{\rm g}(E) \simeq V_{\rm A,i} P_{\rm CR}'(E)/(2 I(k) W_{\rm B})$ \citep{nava:16}, where $V_{\rm A,i}= B/\sqrt{4\pi \rho_{\rm i}}$ is the ion Alfv\'en velocity, $W_{\rm B} = B^2/8\pi$ is the ambient mean magnetic energy density, and $I(k)=(\delta B(k)/B)^2$ is the relative amplitude of perturbations at a wave number $k = k_{\rm res} \sim r_{\rm g}^{-1}$ with respect to the ambient mean magnetic field. The condition $\Gamma_{\rm g} = \Gamma_{\rm in}$ sets the value of $I(k)$. If $I(k) \ll 1$ then parallel and perpendicular CR diffusion coefficients are given by the CR transport quasi-linear theory \citep{schlickeiser:02} by $D_\parallel = 4/3\pi r_{\rm g} v/I(k=r_{\rm g}^{-1})$ and $D_\perp \simeq D_\parallel I(k)^2$, and again $v =\beta_{\rm c} c$ is the particle speed. \\

The parallel diffusion coefficient is then
\begin{equation}
D_\parallel = {4 \over 3\pi}~{r_{\rm g} \beta_{\rm c} c \nu_{\rm in} W_{\rm B} \over V_{\rm A,i} P_{\rm CR}(E)'} \ ,
\end{equation}
or comparing it with $D_{\rm crit}$ given by  Eq.~(\ref{Eq:Dcrit}),
\begin{equation}
{D_\parallel \over D_{\rm crit}} \simeq 4.2~10^3 {E_{\rm GeV} \beta_{\rm c} \nu_{\rm in,-8} (\mu_{\rm i} n_{\rm i,-2})^{1/2} \over P_{\rm CR}'/P_{\rm CR,ref}'} L_{\rm pc}^{-(1+q)} \ .
\end{equation}

We have expressed the CR energy $E_{\rm GeV}$ in GeV units, the ion density $n_{\rm i}$ in units of $10^{-2}~\rm{cm^{-3}}$, the ion-neutral collision frequency in units of $10^{-8}~\rm{s^{-1}}$, and the ion mean mass is $\mu_{\rm i}$. \\

For numerical application we choose the following parameters for WNM and CNM. In the WNM we use $T \simeq 6000$ K, $\mu_{\rm i} =1$, $n_{\rm i,-2} \simeq 0.3,$ and $\nu_{\rm in, -8} \simeq 0.3$. In the CNM we use $\mu_{\rm i}=12$ but provide some estimates for $\mu_{\rm i}=1$ as well, $n_{\rm i,-2} \simeq 2$ and $\nu_{\rm in, -8} \simeq 5 $. In order to have a substantial effect of CR self-generated waves over ISM dynamics, that is to have $D_\parallel \sim D_{\rm crit}$, we need a ratio $P_{\rm CR}'/P_{\rm CR,ref}' > 1$. If we use the conditions adopted in our fiducial case, then $L = 40$ pc and q = 1/3 or q = 1/2. In the WNM we find $P_{\rm CR}'/P_{\rm CR,ref}' \ge 4$ at a CR kinetic energy of 1 GeV in order to get $D_\parallel = D_{\rm crit}$ for q = 1/3 and $P_{\rm CR}'/P_{\rm CR,ref}' \ge 2$ for q = 1/2. In the CNM we find $P_{\rm CR}'/P_{\rm CR,ref}' \ge 390$ still at 1 GeV in order to get $D_\parallel \simeq D_{\rm crit}$ for q = 1/3 and $P_{\rm CR}'/P_{\rm CR,ref}' \ge 210$ for q = 1/2. If we assume $\mu_{\rm i}=1$ we find  $P_{\rm CR}'/P_{\rm CR,ref}' \ge 110$ and $\ge 60$ for q = 1/3 and q = 1/2 respectively. \\

A CR source can perturb the ambient ISM because it can inject some CR overdensity (or overpressure) over scales that can be smaller than typically 100 pc. Hence, the associated CR pressure gradient can exceed $P_{\rm CR, ref}' $ by a factor 10-30 in SNR or even more in star-forming or in starburst galaxies. For instance, \citet{nava:16} consider the propagation of an SNR in warm ISM phases (warm ionised and warm neutral). They investigate the propagation regime of CR release from the forward shock. In that case, the release operates upstream of the shock because high energy CRs have a mean free path larger than a fraction of the shock radius. While escaping ahead of the shock, CRs trigger resonant MHD waves by the streaming instability. The diffusion coefficient is again set by balancing the wave growth rate and the dominant damping process, which is again due to ion-neutral collisions. The authors find that at 10 GeV, so for CRs released late on in the SNR lifetime, diffusion is suppressed\footnote{The suppression is a term often used to describe when the spatial diffusion drops because the scattering frequency of particles by resonantly interacting waves increases. Actually a suppression of the spatial diffusion means that particles are more efficiently scattered.} because of scattering off the self-generated waves down to a factor 100 at time $\sim 10 000$ years after the explosion, while higher CR energies released earlier have a suppression of their diffusion coefficient by a factor 10-30. Calculations in the CNM show similar behaviours although over timescales shorter by a factor of a few kilo-years (Brahimi et al in preparation). However, as far as we know, no calculations have investigated the case of GeV CRs released at the end of shock acceleration or SNR lifetime. To be conservative, at 1 GeV, considering an overpressure of $P_{\rm CR} \simeq 10-30~\rm{eV~cm^{-3}}$ in the ISM over a typical scale-length $R \simeq 10$~pc, typical of a fraction of the radius of a supernova remnant at the end of its lifetime in these phases, we find $P_{\rm CR}'/P_{\rm CR,ref}' \sim 100-300,$ enough with the above requirements to obtain $D_\parallel \simeq D_{\rm crit}$. \\

However, it is mandatory to check that $I(k) \ll 1$ as the streaming instability growth rate has been obtained in the quasi-linear limit. From the condition $\Gamma_{\rm g} = \Gamma_{\rm in}$ we deduce a turbulence level of
\begin{equation}
I(k) \simeq 2.9~10^{-5} \left({(P_{\rm CR}'/P_{\rm CR,ref}') (\mu_{\rm i} n_{\rm i,-2})^{-1/2} \over B_{\rm \mu G} \nu_{\rm in, -8}} \right) \ .
\end{equation}
In the conditions that prevail in partially ionised phases, using $B \simeq 6~\rm{\mu G}$, the ratio $P_{\rm CR}'/P_{\rm CR,ref}' $ 
cannot exceed $\sim 3$ and $\sim 300$ in WNM and CNM respectively for the quasi-linear condition ($I(k) \le 10^{-4}$) to apply. However, this simple calculation made through a rate equilibration likely overestimates the level of magnetic field perturbations produced so the previous limits are likely larger. Non-linear calculations as performed in \citet{nava:16} tend to confirm this assertion. Another caveat of these calculations is that low energy sub-relativistic CRs also released at the end of the acceleration phase or SNR lifetime will ionise the surrounding ISM \citep{padovani:09}. This effect has two implications. Firstly, as the ionisation fraction increases, the ion
density also increases and $D_\parallel/D_{\rm ISM}$ increases as well. This first effect has then a negative feedback on CR ISM dynamical impact. Secondly, as the 
ion density increases, the ion-neutral collision rate decreases and so the damping of self-generated wave also decreases. This second effect has then a positive feedback on the CR ISM dynamical impact. As all quantities above depend on $n_{\rm i}^{1/2}$ , whereas $\nu_{\rm in} \propto n_{\rm n}$ , it is tempting to say that the second effect will take over the first and that the CR confinement and wave generation around SNR will be enhanced. However, an accurate calculation of the propagation of each populations (i.e. low energy sub-relativistic CRs responsible for the ionisation and a mildly relativistic population responsible for the pressure) is needed before stating  any conclusion. The first situation, where ISM ionisation properties are not modified, can still be valid if the mean free path of the sub-relativistic CRs is smaller than the mean free path of mildly relativistic CRs. How these mean free paths compare with each other strongly depends on the nature of the turbulence in the SNR and in its close environment. The latter in practice is strongly dependent on the star formation history that occurs at the location of CR release. \\

A detailed and dynamical calculation of $I(k)$ and $D_\parallel$ with respect to the different types of partially ionised phases will be presented in paper II.  A major difficulty, also discussed in paper II,  in implementing this estimate of CR diffusion in any MHD code is that the scales of the self-generated waves cannot be resolved by the codes. This model of turbulence will be implemented as a sub-grid model.

\subsubsection{CR impact over small star formation scales}
Young stellar objects (YSO) are also potential sites of particle acceleration and can be sources of turbulence injection at ISM scales $\simeq 0.1-1$ pc. \cite{padovani:15} have shown that background CRs cannot be responsible for the observed high ionisation rates  observed in some YSOs (see references therein), which can exceed usual rates by one to four orders of magnitude.
They propose that a process of strong in-situ particle acceleration should occur in these objects. \citet{padovani:16} have identified several potential acceleration sites in YSOs: the jet
termination shocks  in class 0-I YSOs, and accretion shocks in class 0-I or more evolved objects. The region of interaction between the accretion disc and stellar magnetic fields can also produce episodic reconnection events \citep{gouveia:10}. Several groups have investigated the impact of freshly accelerated energetic particles in isolated YSO \citep{rodgers:17, rab:17} or in groups \citep{gaches:18} of YSOs over their surrounding accretion discs or envelopes. These models find that YSOs are, in all these acceleration zones, able to accelerate these CRs in the relativistic regime. These CRs can provide both ionisation and pressure. If ionisation by CRs is an obvious feedback to take into account in models, the feedback produced by CR pressure gradients is a possible supplementary issue to consider in these environments. Of course, in accretion discs, envelopes, or in jets, particle propagation and self-generated wave production are strongly limited by ion-neutral collisions \citep{padovani:16}. On the other hand, CR gradients can develop over scales between 1-1000 AU depending on the site, so leading to higher $P_{\rm ref}'$ values. Among the different acceleration sites invoked above, jets could contribute to drive strong CR gradients. For instance, taking a reference gradient produced by a CR pressure of $1~\rm{eV~cm^{-3}}$ over 1000 AU leads to $P_{\rm CR,ref}' \simeq 1.1\times 10^{-28}~\rm{dyne~cm^{-3}}$. Considering typical molecular cloud core properties as $n_{\rm n} \simeq 10^4~\rm{cm^{-3}}$, $n_{\rm i} \simeq 10^{-2}~\rm{cm^{-3}}$, $\mu_{\rm i} \simeq 29$ \citep{xu:16} and a simulation box $L \simeq 0.1$ pc with q = 1/2, we find $D_{\rm crit} \simeq 10^{22}~\rm{cm^2/s}$ \footnote{We still assume that Larson laws apply at a scale of 0.1-0.01 pc and we use Eq. \ref{Eq:LAR}.} and $D_\parallel \simeq 6.7\times10^{25}~\rm{cm^2/s}$. A dynamical effect of CRs over jet environments seems then to require a very large $P_{\rm CR}'/P_{\rm CR,ref}'$ ratio. This speculative estimate, however, requires a more detailed analysis postponed to a future investigation where the interior of molecular clouds will be considered.

\subsection{Gamma-ray diffuse emission}
Gamma-ray diffuse emission can provide some constraints over CR propagation in different parts of the ISM. Gamma rays are either produced by hadrons through pion production, or by leptons through Inverse Compton emission, or through Bremsstrahlung if the gas density is high enough. The Fermi LAT detector has observed local molecular clouds \citep{Fermi:2012} and giant molecular clouds \citep{Fermi:2012O}, and the Fermi collaboration has obtained a model of diffuse galactic emission \citep{Fermi:2011,Acero:2016}. Among the main results extracted from these data we list: local molecular clouds have gamma-ray spectra compatible with the emission produced by the local CR spectrum observed on Earth but show a variation in normalisation by a factor of 20\%; diffuse gamma-ray emission shows a trend with more intense and harder CR spectra towards small galactic radii to less intense and softer CR spectra in the outer galaxy \citep{Acero:2016}. However, the gamma-ray signal in the outer galaxy is larger than expected from the standard CR diffusion model from known CR source distributions; this is the so-called CR gradient problem. The GeV gamma-ray signal from the Large Magellanic Cloud (LMC) requires a 1-100 GeV CR population with a density of $\sim$30\% larger than the local Galactic value. \\
It is difficult to compare our results with these data. Fermi observed molecular clouds usually in active star formation regions whereas our simulations only consider an ISM free of CR sources. It seems that very active star-forming regions such as the galactic centre or the LMC produce more intense CR components with respect to the local Galactic value. Observations of active CR sources may show that propagation of CRs is different in the environment of these sources. This is for instance the case for the W28 SNR detected at GeV by Fermi \citep{Fermi:2010} and at TeV by HESS \citep{Hess:2008}.  Gamma-ray emission from clouds located near the SNR may be explained if CRs gradually escape from the accelerator, the highest energies being released first. In this framework, diffusion coefficients for TeV CRs required to explain the gamma rays need to be reduced by two orders of magnitude with respect to the standard values \citep{Nava:2013, Yan:2012}.  \citet{Uchiyama:2012} report on the Fermi detection of molecular gas around the W44 SNR. The observations may be interpreted by invoking anisotropic diffusion from the SNR and in that case they would also require some suppression of CR diffusion. These findings advocate in favour of a modified CR propagation in the vicinity of the CR accelerator. The suppression of diffusion is likely a time-dependent process, which depends on the CR energy and the properties of the surrounding ISM \citep[see][]{nava:16}. This transitory aspect makes a comparison between gamma-ray observations and properties of CR-modified ISM turbulence difficult and, at the stage of the present work, premature. The possible impact of CRs over ISM turbulence, in particular around CR sources requires us to include the effect of sources in several ISM numerical realisations and to develop a statistical analysis of the gamma-ray signal from synthetic ISM maps. This modelling is beyond the scope of this study and deserves a future analysis.

\subsection{Model limitation and future works}
In this work, we neglect the effect of CR advection due to the streaming, CR heating due to streaming instability, and CR radiative losses. It has been shown that these processes can have significant effects on large galactic scales for the launching of galactic winds \citep{ruszkowski:17,wiener:17}. It is nevertheless not clear whether they also affect CR propagation at the scales considered in this work. Figures~\ref{Fig:Ma_MC}, \ref{Fig:map_k2}, and \ref{Fig:k2_scatter} indicate that the flow can be sub-Alfv\'{e}nic, in particular within the WNM. As a consequence, we may underestimate the cosmic-ray propagation in the low density regions. These aspects will be investigated in a future work.  

Another limitation comes from the rather idealised set-up used in this study. Although it is more realistic to study ISM dynamics with heating and cooling processes able to handle the thermal instability, the effects of large (galactic) scales and of self-consistent turbulence driving by supernovae or galactic shears are not considered here. It is currently admitted that CR pressure can modify the star formation rate (SFR) by providing extra support, but the details of CR propagation through the galactic scale to the molecular cloud scale are not yet understood. A step forward will be to consider the zooming technique as well as self-gravity,  as proposed in \cite{hennebelle:18} for instance. However, this step is far beyond our current work.

Lastly, we show in Appendix \ref{app:resolution} that our results may not have converged in the trapped CR regime. The CNM gas fraction is underestimated at low resolution and the effect of CR pressure gradients is overestimated. A quantitative analysis has to be done, in particular with respect to the impact the CR can have on the SFR. Nevertheless, we note that CR trapping in the high density regions is robust at high resolution. Local enhancements of the CR energy density are thus still expected, which may have an influence on the ionisation budget of star-forming clouds. In particular, CR ionisation remains the most efficient ionisation process within collapsing dense cores, and appears to control the coupling between the gas and the magnetic fields within the centre of the collapsing region \citep{marchand:16,wurster:18}. Future work is needed in order to quantify the variations of the CR energy density within molecular clouds.  

\section{Conclusion}
\label{section:conclusion}

We have studied the conditions for (GeV) CR trapping within a turbulent and magnetised ISM. First, we have derived an analytical prediction  in Eq.~(\ref{eq:Dcrit_simu}) for the critical diffusion coefficient $D_\mathrm{crit}$ below which CRs are trapped. In the case of a turbulent ISM following the observed scaling relations, we get typical values of $10^{24}-10^{25}$~$\rm{cm^2~s^{-1}}$ for $D_\mathrm{crit}$. Then, we presented numerical experiments of the evolution of a mixture of CRs and gas, which is subject to the thermal instability. We explored the parameter space and varied the initial magnetisation, the initial CR energy density, as well as the turbulence amplitude and driving scale. Our results indicate that our analytical prediction is very robust over the entire explored parameter space. The two regimes of trapped versus free streaming CRs are always recovered, with a transition occurring around $10^{24}-10^{25}$~$\rm{cm^2~s^{-1}}$. The initial CR pressure is found to have a dramatic effect if it exceeds the gas thermal pressure by a factor of ten. In this case, the non-thermal support provided by the CRs is sufficient to inhibit the development of the thermal instability even if the CR diffusion coefficient is larger than the critical value. In the trapped CR regime, the width of the gas density PDF gets narrower, as a result of the larger modified sound speed that accounts for CR pressure. The width of the gas density PDF can be related to the SFR \citep[e.g.][]{hennebelle:11}, and any physical effect, such as turbulence and magnetic fields, can then have an impact on star formation. We have shown that in the trapped CR regime, as well as in CR-pressure-dominated ISM, the width of the PDF is reduced, so that CRs can have a dynamical impact on star formation. Lastly, we discussed possible mechanisms that can lower the diffusion coefficient to values that are smaller than the canonical admitted ones of a few $10^{28}$~$\rm{cm^2 s^{-1}}$. The turbulence generation by CR streaming especially around CR sources, as well as models for large-scale turbulence, appear to be potential mechanisms able to modify diffusion coefficient values. \\

To summarise, we have shown that CR pressure gradients can develop within the ISM forming molecular clouds, if the condition for CR trapping is satisfied or if the CR pressure largely exceeds the gas thermal pressure. CR pressure gradients affect the gas dynamics, as well as the propagation of CRs. This feedback loop needs to be better understood in future works, which will account for local variations of the CR diffusion coefficient as a function of large-scale turbulent models, as well as sub-grid models for turbulence generated by CR streaming.

\begin{acknowledgements}
We thank the anonymous referee for his or her constructive comments that helped to improve the quality of the manuscript. BC thanks Andrew McLeod for his contribution to the driven turbulence module. This work was supported by the CNRS programmes ``Programme National de Cosmologie et Galaxies'' (PNCG), ``Physique et Chimie du Milieu Interstellaire'' (PCMI),  and ``Programme national des hautes \'energies'' (PNHE). This project was supported by the IDEXLyon project (contract n°ANR-16-IDEX-0005) under the auspices of the University of Lyon. This work was granted access to the HPC resources of CINES (Occigen) under the allocation DARI A0020407247 made by GENCI.  Some of the figures were created using the \href{https://bitbucket.org/nvaytet/osiris}{\texttt{OSIRIS}}\footnote{\url{https://bitbucket.org/nvaytet/osiris}} visualisation package for \texttt{RAMSES}.
\end{acknowledgements}

\bibliographystyle{aa}
\bibliography{biblio}

\appendix
\section{Convergence with the resolution\label{app:resolution}}

\begin{figure*}[thb]
        \includegraphics[width=1\textwidth]{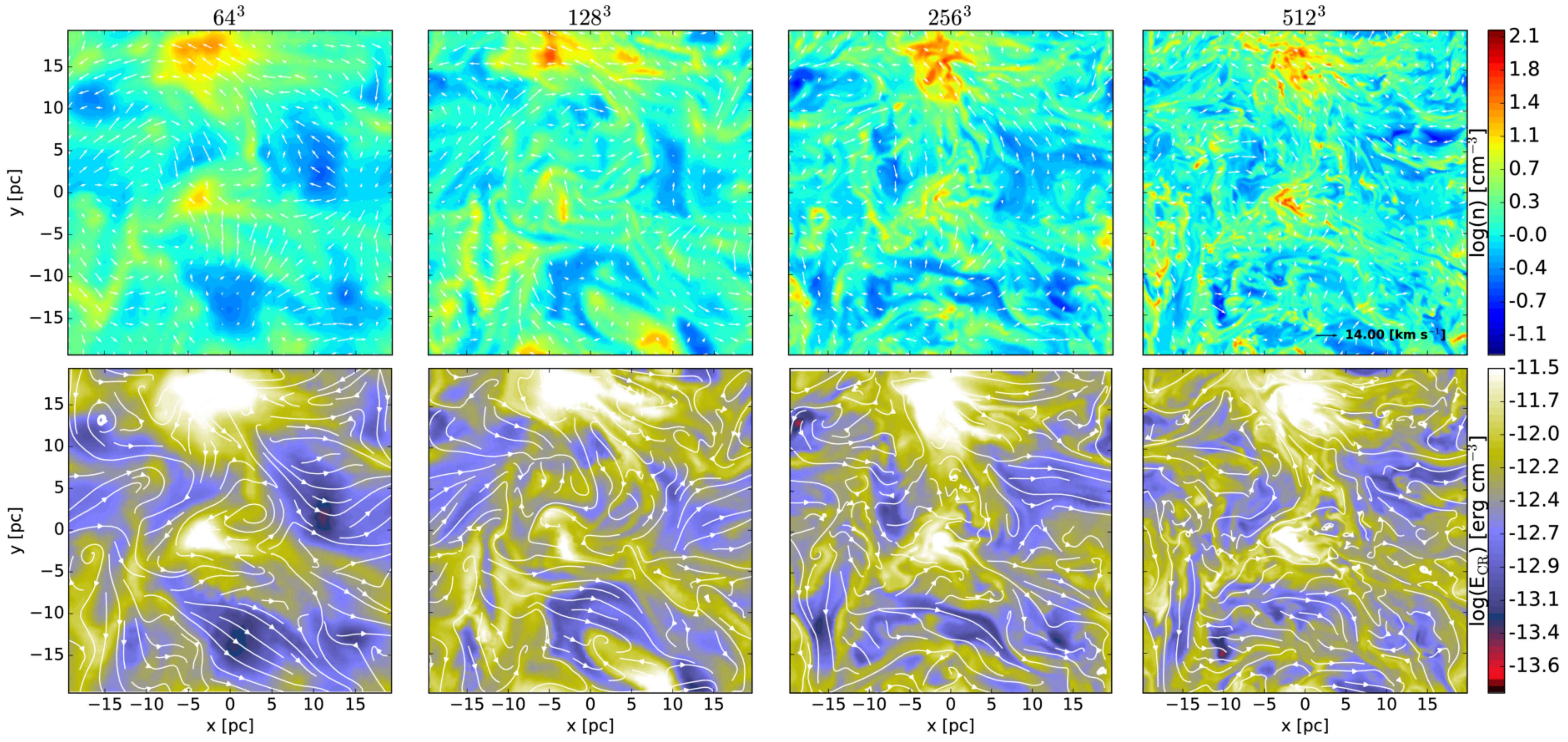}
        \caption{Gas density (top) and CR energy density (bottom) maps  for the fiducial runs with $D_0=10^{24}\, \rm cm^2 \, s^{-1}$ in the plane $z=20$~pc at a time corresponding to $1.2t_\mathrm{cross}$ and a resolution of $64^3$, $128^3$, $256^3$, and $512^3$ (from left to right).}
        \label{Fig:D24_res}
\end{figure*}

In this appendix, we perform a resolution study for the fiducial case, with $k_\mathrm{turb}=2$, and for $D_0=10^{24}$~and $D_0=10^{26}$~cm$^2$~s$^{-1}$. We perform four calculations with $D_0=10^{24}$~cm$^2$~s$^{-1}$ and a resolution ranging from $64^3$ to $512^3$, and three calculations with $D_0=10^{26}$~cm$^2$~s$^{-1}$ and a resolution ranging from $64^3$ to $256^3$. We do not present results at a resolution of $512^3$ with  $D_0=10^{26}$~cm$^2$~s$^{-1}$ because the CPU time needed to reach only one dynamical time is too expensive. This comes from the bad efficiency of our implicit solver when dealing with rather smooth CR energy density and highly anisotropic diffusion.

Figure \ref{Fig:D24_res} shows gas density and CR energy density maps for the four calculations done with $D_0=10^{24}$~cm$^2$~s$^{-1}$. The denser structures get more compact as resolution increases. We attribute this behaviour to the poor sampling of the CNM structures with a resolution less than $256^3$ for a 40~pc box. \cite{audit:05} argued that the typical size of the fragments of CNM is $\simeq 0.1$~pc. We are clearly sub-sampling this scale with our fiducial resolution, which indicates that we cannot properly describe the structure of the densest gas, that is, the molecular clouds.  Nevertheless, the cooling length in the WNM,  $\simeq 1-3$~pc at a density of 2~cm$^{-3}$ \citep{hennebelle:99}, is always resolved. The high CR energy density regions also exhibit smaller structures as resolution increases, but not as much as the gas density ones. 

\begin{figure}[t]
        \includegraphics[width=0.5\textwidth]{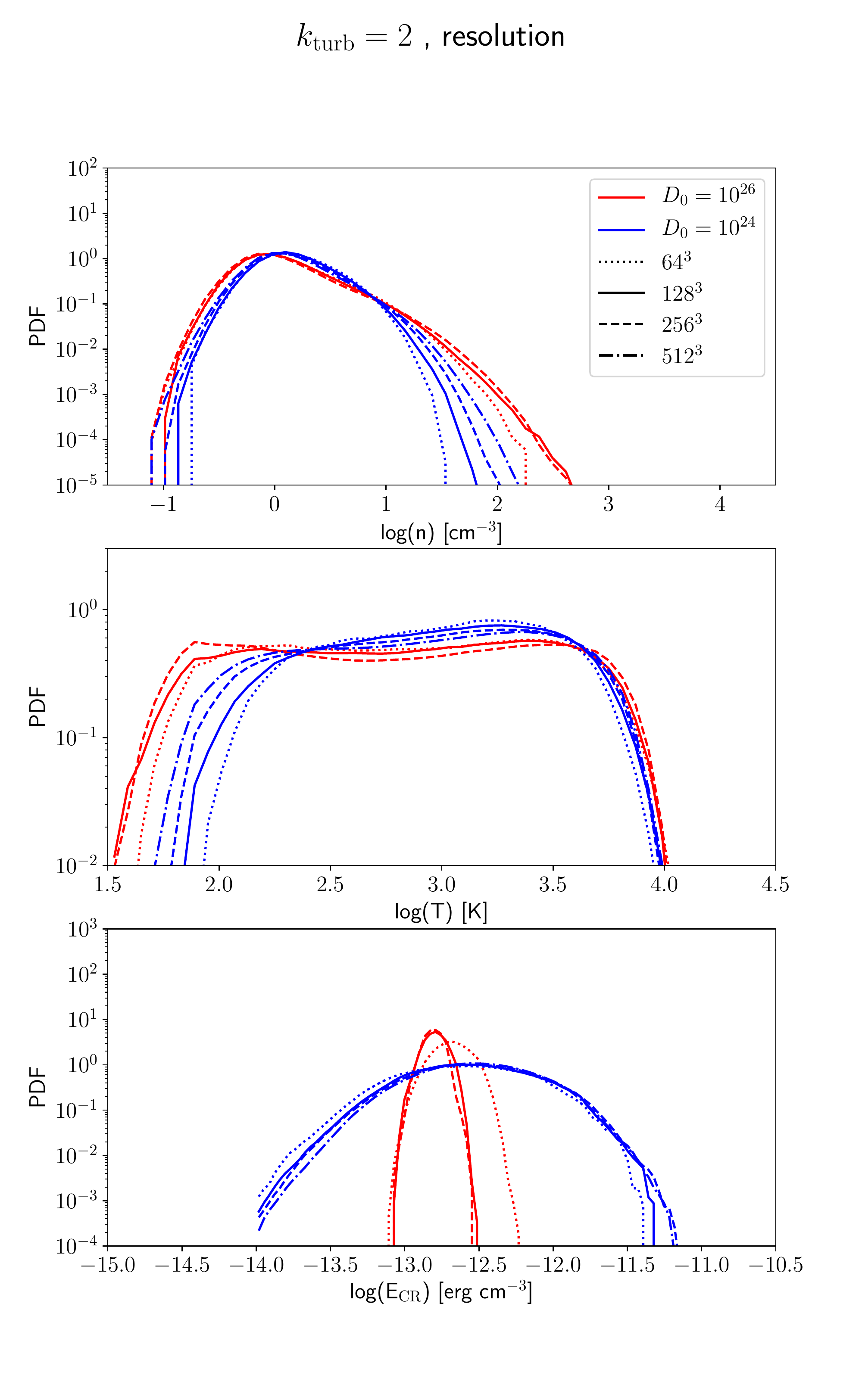}
        \caption{Same as Fig.~\ref{Fig:pdf_k2} for different numerical resolutions in the fiducial case ($k_\mathrm{turb}=2$)  with $D_0=10^{24}\, \rm cm^2 \, s^{-1}$ (blue) and $D_0=10^{26}\, \rm cm^2 \, s^{-1}$ (red), but at times greater than 1~$t_\mathrm{cross}$.}
        \label{Fig:pdf_res}
\end{figure}

Figure \ref{Fig:pdf_res} shows the PDFs of the gas density, temperature, and CR energy density of the seven different calculations at times greater than 1~$t_\mathrm{cross}$. The gas density and temperature PDFs broaden as resolution increases. In particular, the PDFs of the case with $D_0=10^{24}$~cm$^2$~s$^{-1}$ do not show convergence as resolution increases. The gas fraction in the CNM increases significantly in the $512^3$ run. This feature was also reported in other works that report more pronounced phase transition at higher resolution \citep{seifried:11}. Nevertheless, the CR energy density PDFs are not very sensitive to the resolution, except for the high energy tail. Importantly, the qualitative behaviour of the CR energy density does not change if resolution increases, with a transition to the trapped CRs regime with a low diffusion coefficient. We note, however, that given the fiducial resolution we use of $128^3$, our study cannot be quantitative for the CNM gas fraction and structures.

\section{Isotrop versus anisotrop CR diffusion\label{appendix:isotropic}}
\begin{figure}[t]
  \includegraphics[width=0.5\textwidth]{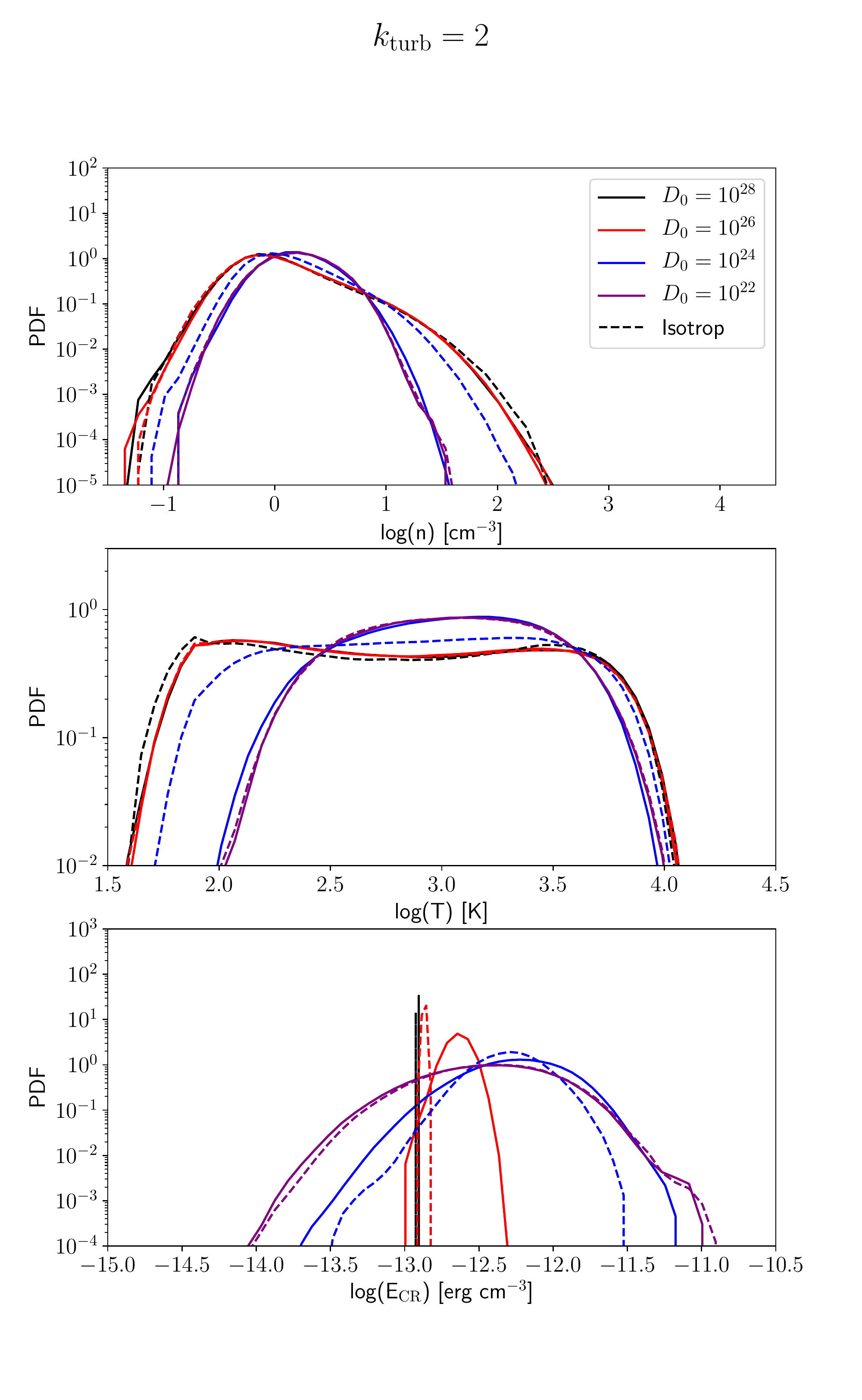}
  \caption{Same as Fig.~\ref{Fig:pdf_k2}. Effect of anisotropic diffusion versus isotropic for the fiducial runs.}
    \label{Fig:pdf_iso}
\end{figure}

Cosmic rays are charged particles and their propagation is therefore highly dependent on the magnetic field configuration, which is thus anisotropic in nature. We show in this appendix to what extent the anisotropic case differs from a simple isotropic diffusion case. We perform four runs of the fiducial case, with a diffusion coefficient ranging from $10^{22}$ to $10^{28}$~cm$^2$~s$^{-1}$ , but we set the perpendicular diffusion coefficient equal to the parallel one. 

Figure~\ref{Fig:pdf_iso} compares the PDFs of the gas density, gas temperature, and CR energy density for the cases with anisotropic diffusion and with isotropic diffusion. For the two extreme cases, $D_0= 10^{22}$ and $D_0= 10^{28}\, \rm cm^2 \, s^{-1}$, the PDFs are not very sensitive to the nature of the diffusion process. In the first case, the diffusion coefficient is too low and CRs are trapped anyway, while in the second case diffusion is too fast. 
We find differences in the CR energy density PDF with $D_0= 10^{26}\, \rm cm^2 \, s^{-1}$. The PDF is indeed narrower with isotropic diffusion because the diffusion process is more efficient. With $D_0= 10^{24}\, \rm cm^2 \, s^{-1}$, the three PDFs are different if isotropic diffusion is used. The CRs can escape more efficiently and cannot build up pressure gradients that act on the gas motions. Since this value is close to the critical value, a slight change of roughly a factor~of three in the overall diffusion coefficient makes important differences. 
Overall, these tests indicate that anisotropic diffusion has to be taken into account if one aims at describing properly the transition to CR trapping. 

\end{document}